\documentclass[12pt]{article}
\usepackage{latexsym}
\usepackage{epsfig,amssymb,euscript}
\usepackage{amsmath}
\usepackage{array,calc,epsfig}
\usepackage{citesort}

\def\be{\begin{equation}}
\def\ee{\end{equation}}
\def\bea                                                             {\begin{eqnarray}}
\def\eea{\end{eqnarray}}

\numberwithin{equation}{section} 
\usepackage[]{graphicx}

\def\cald         {{\cal D}}

\def\calf         {{\cal F}}

\def\call         {{\cal L}}

\def\caln         {{\cal N}}

\def\calu         {{\cal U}}

\def\calw         {{\cal W}}

\def\Re           {{\rm Re\hskip0.1em}}
\def\Im           {{\rm Im\hskip0.1em}}

\def\sqr#1#2{{\vcenter{\vbox{\hrule height.#2pt
 \hbox{\vrule width.#2pt height#1pt \kern#1pt \vrule width.#2pt}\hrule
 height.#2pt}}}}

\def\slashchar#1{\setbox0=\hbox{$#1$}           
\dimen0=\wd0                                 
\setbox1=\hbox{/} \dimen1=\wd1               
\ifdim\dimen0>\dimen1                        
\rlap{\hbox to \dimen0{\hfil/\hfil}}      
#1                                        
\else                                        
\rlap{\hbox to \dimen1{\hfil$#1$\hfil}}   
/                                         
\fi}

\begin{document}
\font\cmss=cmss10 \font\cmsss=cmss10 at 7pt
\leftline{\tt hep-th/yymmnnn}

\vskip -0.5cm
\rightline{\small{\tt KUL-TF-06/02}}

\vskip .7 cm

\hfill
\vspace{18pt}
\begin{center}
{\Large \textbf{D-branes on general ${\cal N}=1$ backgrounds:\\ superpotentials and D-terms}}
\end{center}

\vspace{6pt}
\begin{center}
{\large\textsl{Luca Martucci}}

\vspace{25pt}
\textit{\small  Institute for Theoretical Physics, K.U. Leuven,\\ Celestijnenlaan 200D, B-3001 Leuven, Belgium}\\  \vspace{6pt}
\end{center}

\vspace{20pt}

\begin{center}
\textbf{Abstract}

\end{center}

\vspace{4pt} {\small \noindent We study the dynamics governing space-time filling D-branes on  Type II flux backgrounds preserving four-dimensional ${\cal N}=1$ supersymmetry. The four-dimensional superpotentials and D-terms are derived. The analysis is kept on completely general grounds thanks to the use of recently proposed generalized calibrations, which also allow one to show the direct link of the superpotentials and D-terms with BPS domain walls and cosmic strings respectively. In particular, our D-brane setting reproduces the tension of  D-term strings found  from purely four-dimensional analysis. The holomorphicity of the superpotentials is also studied and a moment map associated to the D-terms is proposed. Among different examples, we discuss an application to the study  of D7-branes on $SU(3)$-structure backgrounds, which reproduces and generalizes some previous results.}

\vspace{4cm}


\vfill
\vskip 5.mm
\hrule width 5.cm
\vskip 2.mm
{\small
\noindent e-mail: luca.martucci@fys.kuleuven.be}

\newpage

\tableofcontents
\bigskip
\bigskip
\section{Introduction}

The study of string theory compactifications to four dimensions with non-trivial fluxes is not only interesting by itself, but seems to be necessary if we hope to use string theory to describe realistic scenarios. Moreover,  backgrounds with fluxes arise naturally also in the  context of the gauge-gravity correspondence.   D-branes play a central role in several aspects of these constructions and thus the study of their properties on nontrivial flux backgrounds is of both formal and phenomenological interest.

In this paper we study the dynamics of  space-time filling
D-branes in the most general Type II backgrounds preserving four-dimensional 
Poincar\'e invariance and ${\cal N}=1$ supersymmetry.
The aim is that of presenting a unified analysis that automatically
includes a large class of cases, 
having ${\cal N}=1$ background supersymmetry as the unique requirement. This analysis obviously
includes as special subcases the ${\cal N}=2$ backgrounds that are
obtained by turning off the Ramond-Ramond (RR) fields, and in
particular the limit in which the internal space reduces to a
standard Calabi-Yau space.

D-brane dynamics in $\caln=2$ compactifications on standard
Calabi-Yau three-folds constitute an active past and present
topic of research  (for  reviews and  complete lists of
references see for example \cite{doug,aspinwall}). One approach, 
that we will  follow in the present paper, is to consider
D-branes filling the four flat directions and wrapping some internal
cycle, describing the system by an effective  four-dimensional
$\caln=1$  theory.  The well-known geometrical properties of the
underlying Calabi-Yau spaces allow one to employ a series of familiar
technics. Many problems can be addressed systematically by using
the two integrable structures of the Calabi-Yau, i.e. the complex
and symplectic structures, and for example the associated twisted
topological theory constitutes an efficient way to inspect the
holomorphic sector of the theory \cite{witten92}.

In general, the reduction of the background supersymmetry to  $\caln=1$, obtained by giving non-trivial expectation value to the internal fluxes, can drastically change the geometry of the internal space\footnote{For a  review on flux 
compactifications see for example \cite{granarev}.}. In particular, the
symplectic and complex  structures cease to be defined
 in general and, even  in  cases when they are both defined, they may not be
simultaneously integrable. However, as discussed in \cite{gmpt2}
for a wide class of $\caln=1$ vacua, the minimal supersymmetry still
imposes an integrable structure on the internal
manifold that can be described as a generalized complex structure by using the language of generalized complex geometry \cite{hitchin,gualtieri}. The complex and
symplectic structures are substituted by a pair of pure spinors
(that are particular kinds of polyforms - formal sums of forms of different degrees) of definite and opposite
parity, that are associated to corresponding generalized {\em almost}
complex structures. The background supersymmetry conditions are
written in terms of these two pure spinors and imply that one of the
associated generalized almost complex structures is actually integrable, while the RR
background fluxes provide an obstruction to the integrability of the
other.

In this paper we will consider the most general class of $\caln=1$
backgrounds  admitting  supersymmetric static D-brane
configurations. These backgrounds constitute a subclass of the vacua
analyzed in \cite{gmpt2} and we will refer to them as {\em
D-calibrated} backgrounds. The name is justified by the fact that,
as shown in \cite{lucal}, these supersymmetric backgrounds can be
completely characterized  in terms of a new kind of generalized
calibrations associated to the possible supersymmetric static
D-brane configurations (i.e. filling two, three or all four
space-time directions)\footnote{Similar generalized calibrations
were introduced  in \cite{koerber} for the subclass of ${\cal N}=2$
backgrounds obtained by switching off the RR fields.}.  The
generalized calibrations are essentially given by the real and
imaginary parts of the background pure spinors, and provide an elegant physical 
interpretation for them.

Introducing a space-time filling D-brane wrapping some internal generalized cycle (defined as cycle with a world-volume field-strength on it) on these D-calibrated backgrounds, the effective four-dimensional
description should admit an  $\caln=1$ structure. Indeed, we will show how
it is  possible to identify superpotentials and  D-terms that
can be written in a completely general form in terms of  the
underlying generalized calibrations  (and then of the background
pure spinors). The associated  F-flatness and D-flatness conditions
are equivalent to the supersymmetry/calibration condition found in
\cite{lucal}.

Regarding the  superpotential, we will see how it only involves
the background {\em integrable} pure spinor and the associated
F-flatness condition requires that the D-brane must wrap a {\em
generalized  complex submanifold}, as defined in \cite{gualtieri}.
This  result can be seen as an extension of the ``decoupling
statement" of \cite{quintic}, that  in the present context  can be
rephrased as the requirement that the superpotential only `sees' 
the underlying (integrable) generalized complex structure. The
superpotentials we find may be adopted for the topological branes 
\cite{Kapustin,zabzine,Zucchini,Li,Kapustin:2005vs}
of the associated topological models \cite{Kapustin,kapu2,gmpt2}.
Our superpotentials generalize known superpotentials for D-branes on
Calabi-Yau manifolds, as studied  for example in
\cite{witten92,wittenQCD,kachru,agana}. They are also in agreement with
previous results for D7-branes with world-volume and/or background
fluxes  \cite{lust2,louis2}. We will discuss the holomorphic
properties of the superpotentials and shall see how they can  be
addressed in a unified way, again generalizing  previous results
for D-branes on Calabi-Yau spaces (see for example the discussion of
\cite{thomas,hori}).

It is well known that the tension of a possible BPS domain wall in an
$\caln=1$ theory is expressed uniquely in term of the
superpotential. This relation has been used for example in
\cite{gukov1,gukov2} for deriving flux induced superpotentials for
the closed string moduli. Using the underlying generalized
calibrations, we will see how the same approach can also be used to
give an alternative and more physical derivation of the D-brane
superpotentials, thus obtaining  a non-trivial consistency
check of our results.

Regarding the D-flatness condition, in the standard Calabi-Yau
case, it can be seen as a deformed Hermitian-Yang-Mills equation for
the holomorphic connection on the holomorphic B-cycles, while for
Lagrangian A-branes it corresponds to the additional ``speciality"
conditions (a discussion and more references can be found in \cite{doug}). 
These conditions are equivalent  to the vanishing of a moment map associated to the $U(1)$ gauge
symmetry on the D-brane through an appropriate symplectic structure on
the configuration space \cite{thomas,hori}. The vanishing of the
moment map  provides a transversal slicing for the imaginary
extension of the gauge group action, whose complexification is a
symmetry of the superpotential. An extension of this approach to the
case of $SU(3)$-structure backgrounds  has been discussed in
\cite{Grange}. We will propose a symplectic form that  generalizes
 the known  symplectic structures  of the
above mentioned particular subcases to our more general setting.
 Using this, the $U(1)$ gauge
symmetry on the wrapped cycle is associated to a moment map whose
vanishing condition is equivalent the our general D-flatness
condition.

The insight given by  the generalized calibrations
characterizing these backgrounds allows us to derive another
interesting physical result. Namely, the D-term turns out to be
strictly related to the  BPS cosmic strings obtained by wrapping D-branes filling only two flat
space-time directions
around internal generalized cycles. First, we will discuss how the D-flatness
condition can be satisfied only if a certain topological constant
$\xi$ vanishes. This constant  can be identified with the
Fayet-Iliopoulos term in the effective four-dimensional description.
Then, we will see how the BPS cosmic string tension computed from
our D-brane setting matches precisely the BPS cosmic string tension
$T_{\rm string}=2\pi\xi$ obtained from $\caln=1$ supergravity
\cite{bbs,jose,toinestring}, which should describe a D$\bar{\rm D}$-brane
pair. This result provides a non-trivial check  of the
identification proposed in \cite{toinestring} between these D-term
supergravity string solitons and the effective cosmic strings
obtained by wrapping D-branes.

The plan of the paper is as follows.   In section \ref{sec1} we review the basic results of \cite{lucal}, i.e. the D-brane supersymmetry conditions and the associated generalized calibrations. In section \ref{sec2} we show how, starting from the Dirac-Born-Infeld (DBI) plus Chern-Simons (CS) action for D-branes, we can organize the four-dimensional potential in an explicit ${\cal N}=1$ form, recognizing the supersymmetry/calibration conditions in the  F- and D-flatness conditions of the four-dimensional description. In section \ref{supsec} we introduce a superpotential that gives rise to the  F-flatness condition. This can be written in a universal way by using the underlying integrable pure spinor. In section \ref{supDW} we give an alternative  derivation of the superpotential by using domain wall D-brane configurations. Cosmic string D-brane configurations are  considered in section \ref{Dstrings}, stressing their relation with the Fayet-Iliopoulos contribution to the D-term and giving a general nontrivial argument in favor of their identification with the supergravity cosmic strings constructed in \cite{toinestring}.  The holomorphicity properties of the superpotentials  are studied in section \ref{hol}, where an almost complex structure is introduced on the  D-brane configuration space from the $SU(3)\times SU(3)$ structure of the underlying background.  In section  \ref{complex} we discuss the reduction of this almost complex structure to the superpotential critical subspace. In section \ref{symplectic} we turn to the D-flatness condition and see how it can be interpreted as the vanishing of a moment map associated to the world-volume gauge symmetry by an appropriately defined symplectic structure. Section \ref{examples} is dedicated to some explicit examples in the more specific case of backgrounds with internal $SU(3)$-structure \cite{gmpt}: we will consider D3, D5, D6 and D7-branes, with particular attention being paid to the last case, for which our general analysis reproduces  results present in the literature (see for example \cite{softuranga,louis1,lust3,lust2,louis2,marchesano}). Finally, in section \ref{discussion} we present our concluding remarks.  Appendix \ref{appA} contains a more detailed discussion on our parametrization of the infinitesimal  deformations of  generalized cycles.

\section{D-calibrated ${\cal N}=1$ vacua}
\label{sec1}

In this paper we consider the most general Type II ${\cal N}=1$ backgrounds with four-dimensional
Poincar\'e  invariance which admit possible supersymmetric D-branes filling one or more flat space directions
and wrapping some internal cycle. As discussed  in \cite{lucal}, all the backgrounds satisfying these conditions
consists of a subclass of the family of ${\cal N}=1$ vacua analyzed in \cite{gmpt2}, that we refer to as {\em D-calibrated} since they can be completely characterized by the existence a kind of  generalized calibrations \cite{lucal} as we are going to review in this section.

Let us discuss briefly the  main properties of the ${\cal N}=1$ D-calibrated backgrounds,
following the conventions of \cite{lucal}. The ten dimensional metric can be written in the general form
\bea
ds^2=e^{2A(y)}dx^\mu dx_\mu+ g_{mn}(y)dy^m dy^n\ ,
\eea
where $x^\mu$, $\mu=0,\ldots,3$ label the four-dimensional flat space, and $y^m$, $m=1,\ldots,6,$ the internal space.
The $B$-form field-strength $H=dB$ can have legs only along internal directions, while the generalized  RR field-strengths
\bea
F_{(n+1)}=dC_{(n)}+H\wedge C_{(n-2)}\ ,
\eea
are allowed to have the restricted  form
\bea                                                                                                                          \label{RRdec}
 F_{(n)}=\hat F_{(n)}+Vol_{(4)}\wedge \tilde F_{(n-4)}\ .
\eea All the fields appearing in this ansatz (including  the dilaton $\Phi$) can depend only on the internal coordinates $y^m$.
Note also that the usual electric-magnetic Hodge duality relating lower and higher degree RR field-strength
translates into the relation $\tilde
F_{(n)}=(-)^{\frac{(n-1)(n-2)}{2}}*_6 \hat F_{(6-n)}$ between
their internal components.

Starting from this bosonic ansatz, the ${\cal N}=1$ supersymmetry imposes that there exist four independent ten dimensional Killing spinors that can be written in terms of an arbitrary four-dimensional constant spinor $\zeta_+$ of positive chirality and two internal six-dimensional spinors $\eta^{(1)}$ and $\eta^{(2)}$.  The resulting Killing equations give strong constraints on the background bosonic ansatz.  The important result proved in \cite{gmpt2}  is that these supersymmetry constraints on the background fields
can be nicely written  in terms of the following two polyforms of definite parity
\bea
\Psi^+=\sum_{k\geq 0}\Psi^+_{(2k)}\quad\quad \Psi^-=\sum_{k\geq 0}\Psi^-_{(2k+1)} ,
\eea
corresponding via the usual Clifford map to the bispinors\footnote{See \cite{Witt} for a previous analysis  using these bispinors in pure NS backgrounds.}
\bea
\slashchar{\Psi}^+=\eta^{(1)}_+\otimes\eta^{(2)\dagger}_{+}\quad,\quad \slashchar{\Psi}^-=\eta^{(1)}_+\otimes\eta^{(2)\dagger}_{-}\ .
\eea
$\Psi^\pm$ can be seen to be pure spinors in the context of the generalized complex geometry and define corresponding generalized almost complex structures\footnote{More explicit expressions for the pure spinors $\Psi^\pm$ can be found in \cite{gmpt2}. The case of D-calibrated $SU(3)$-structure vacua, which include the standard $\caln=2$  compactifications on flux-less Calabi-Yau's as a subcase, will be discussed in detail in section \ref{examples}.}. As we will presently recall, in the ${\cal N}=1$ vacua we are interested in, only one of these two generalized almost complex structure will actually be integrable \cite{gmpt2}.

Note  that not all the possible ${\cal N}=1$ solutions with 4d Poincar\'e invariance  can be studied in these terms. Indeed there are some pure NS ${\cal N}=1$ vacua
\cite{hull,strominger,gauntlett} that cannot be incorporated in these class of backgrounds, since in these cases one of the two internal spinors vanishes and then both $\Psi^\pm$ vanish as well,  spoiling of any mean the above approach. However, as discussed in \cite{lucal}, the only ${\cal N}=1$ backgrounds that can admit supersymmetric (static) D-branes filling one or more flat space directions are those whose internal spinors have the same norm
\bea                                                                                                                          \label{normcond}
||\eta^{(1)}||^2=||\eta^{(2)}||^2=|a|^2\ .
\eea
This means that the cases we are interested in can be completely covered by the description given in \cite{gmpt2} and the condition (\ref{normcond}) allows also to characterize this class of ${\cal N}=1$ backgrounds as D-calibrated.
More explicitly, taking into account the additional requirement given in (\ref{normcond}), the supersymmetry conditions
for the backgrounds can be split in two parts.
One relates the warp factor $A$ to the norm of the internal spinor
\bea
d|a|^2=|a|^2dA&\Rightarrow& |a|^2=ce^A\ ,
\eea
for some constant $c$. This relation is a direct consequence of the 4d ${\cal N}=1$ supersymmetry as it is equivalent to require that $\bar\epsilon \Gamma^\mu\epsilon$ (here $\epsilon$ is the 10d Killing spinor doublet) must be a (constant) Killing vector generating the 4d spacetime translations.

The other supersymmetry conditions involve the two pure spinors $\Psi^\pm$ characterizing our backgrounds\footnote{See \cite{lucal} for the conventions we are using and how they are related
to the ones adopted in \cite{gmpt2}.}
 \bea                                                                                                                          \label{backsusy} e^{-2A+\Phi}d_H\big(
e^{2A-\Phi}\Psi_1\big)&=&dA\wedge \bar \Psi_1 +\frac{i|a|^2}{8}e^\Phi\tilde F\ ,\cr
d_H\big(e^{2A-\Phi}\Psi_2\big)&=&0\ , \eea where \bea d_H=d+H\wedge\eea is the $H$-twisted differential (such that $d_H^2=0$) and  for Type IIA we have \bea                                                                                                                          \label{defA}
&&\Psi_1=\Psi^-\quad,\quad \Psi_2=\Psi^+\quad {\rm and}\quad \tilde F=\tilde F_A=\tilde F_{(0)}+\tilde F_{(2)}+\tilde F_{(4)}+\tilde F_{(6)}\ , \eea while for Type IIB \bea                                                                                                                          \label{defB} &&\Psi_1=\Psi^+\quad, \quad \Psi_2=\Psi^-\quad {\rm and}\quad
\tilde F=\tilde F_B=\tilde F_{(1)}+\tilde F_{(3)}+\tilde F_{(5)}\ . \eea

As proved in \cite{lucal}, the equations given in (\ref{backsusy}) can be completely characterized in terms of properly defined generalized calibrations associated to possible static supersymmetric D-branes. These are given by D-branes filling two (strings), three (domain walls) or all four flat space-time directions, and wrapping some internal generalized cycle $(\Sigma,\calf)$, i.e. a cycle  $\Sigma$ with a possible general world-volume field strength $\calf$ on it (which satisfies the Bianchi identity $d\calf=P_\Sigma[H]$). The associated generalized calibrations are given by  $d_H$-closed formal sums of (real) forms $\omega=\sum_k \omega_{(k)}$ of definite degree parity   which are written in terms of the pure spinors $\Psi^{\pm}$ and are properly energy minimizing  when combined with the world-volume field strength $\calf$. More explicitly, for any generalized cycle $(\Sigma,\calf)$,
\bea                                                                                                                          \label{algcond}
P_\Sigma[\omega]\wedge e^\calf|_{\rm top}\leq {\cal E}(\Sigma,\calf)\ ,
\eea
where ${\cal E}(\Sigma,\calf)$ refers to the energy density of the D-brane wrapping $(\Sigma,\calf)$. Note that, if on one hand the form of the generalized calibration is completely general for  wrapped cycles of any dimension, on the other hand it {\em does}  depend explicitly on the number of flat space-like directions  filled by the D-brane\footnote{This effect comes directly from the $\caln=1$ supersymmetry of the background (that, for example, implies a nontrivial warp factor). In the $\caln=2$ limit reached by turning off the RR fields, the form of the generalized calibrations acquires an arbitrary phase \cite{koerber} and does not depend on the filled flat directions.}.   If we introduce the normalized pure spinors
\bea                                                                                                                          \label{normps}
\hat\Psi^{\pm}=-\frac{8i}{|a|^2}\Psi^\pm\ ,
\eea the generalized calibrations for space-time filling D-branes are given by \cite{lucal}
\bea                                                                                                                          \label{4dcal}
\omega^{\rm (4d)}=e^{4A}\Big[ e^{-\Phi}{\rm Re}\hat\Psi_1-\tilde C\Big]\ ,
\eea
where $\tilde C=\sum_{k}\tilde C_{(k)}$ with $k$ even in Type IIB and odd in Type IIA, and $\tilde C_{(k)}$ are potentials for $\tilde F_{(k+1)}$, such that $\tilde F_{(k+1)}=d_H \tilde C_{(k)}+4dA\wedge \tilde C_{(k)}$.  The generalized calibrations for four-dimensional strings and domain walls are given by
\bea                                                                                                                          \label{others}
\omega^{\rm(string)}=e^{2A-\Phi}{\rm Im}\hat\Psi_1\quad,\quad \omega^{\rm(DW)}=e^{3A-\Phi}{\rm Re}(e^{i\theta}\hat\Psi_2)\ .
\eea
The calibration for the domain walls contain an a priori arbitrary phase specifying the preserved half of the four-dimensional ${\cal N}=1$ supersymmetry.  The inequality (\ref{algcond}) for each of the generalized calibrations comes from completely algebraic considerations while the differential requirement that they are  $d_H$-closed is completely equivalent to the background supersymmetry conditions given in (\ref{backsusy}).   Note also  that the generalized calibrations for the space-time filling and four-dimensional string D-branes involve the {\em non-integrable} pure spinor, while the generalized  calibration for four-dimensional domain wall D-branes involves the {\em integrable} pure spinor.

A supersymmetric D-brane configuration can be completely characterized as a D-brane wrapping a {\em generalized calibrated cycle}, i.e. a generalized cycle $(\Sigma,\calf)$ which saturates in each point the upper bound in (\ref{algcond}). As discussed in \cite{lucal}, this condition can be split in an equivalent pair of conditions. In the case of the space-time filling D-branes (on which we focus from now on), these are given by\footnote{Calibrated  strings are alternatively defined by the conditions $P_\Sigma[\Re \hat\Psi_1]\wedge e^\calf|_{\text{top}}=0$ and $P_\Sigma[(\imath_m+g_{mn}d y^n\wedge)\hat\Psi_2]\wedge e^\calf|_{\text{top}}=0$, while domain walls by $P_\Sigma[\Im (e^{i\theta}\hat\Psi_2)]\wedge e^\calf|_{\text{top}}=0$ and $P_\Sigma[(\imath_m+g_{mn}d y^n\wedge)\hat\Psi_1]\wedge e^\calf|_{\text{top}}=0$.}

\bea                                                                                                                          
\label{FDterms}
&& P[dy^m\wedge \hat\Psi_2 +g^{mn}\imath_n \hat\Psi_2]\wedge e^\calf|_{\rm top}=0\quad,\quad\quad\, {\rm F-flatness}\ ,\cr
&& P[{\rm Im}\hat\Psi_1]\wedge e^\calf|_{\rm top}=0\quad,\quad\quad\quad\quad \quad\quad\quad\quad\ \ {\rm D-flatness}\ .
\eea
The reason why we have used the names F-flatness and D-flatness will be the focus of the following discussions. For the moment, let us only recall that the F-flatness imposes that the D-brane must wrap a generalized complex submanifold \cite{lucal} and then specifies the generalized complex geometry of the supersymmetric D-branes. In the $SU(3)$-structure cases, where the internal manifold is either complex (IIB) or symplectic (IIA), this requirement is completely equivalent \cite{gualtieri} to require that the D-brane must be holomorphically embedded with $\calf_{(2,0)}=0$ in Type IIB, and must wrap Lagrangian or more general coisotropic \cite{kapu} generalized cycles in Type IIA.  This is completely analogous to what happens in the flux-less Calabi-Yau case, where the above geometrical conditions must be supplemented by a stability condition which can be read as a deformed Hermitian-Yang-Mills equation for B branes and the ``speciality" condition for Lagrangian A-branes and their coisotropic generalization \cite{mmms,kapuli}. As discussed in a series of paper by Douglas and collaborators (see e.g. \cite{doug} for a review), this stability condition can be seen as a D-flatness condition, obtained by imposing the vanishing of an associated moment map. In the following sections we will discuss in detail the above F-flatness and D-flatness in our general setting considering ${\cal N}=1$ backgrounds, trying to clarify their meaning and their relation with the results already known in the Calabi-Yau case.

\section{The four-dimensional point of view}
\label{sec2}

In this section, using the results of \cite{lucal} reviewed in section \ref{sec1}, we would like to pass to a four-dimensional description of the dynamics of the space-time filling D-branes, which should be ultimately described by a four-dimensional ${\cal N}=1$ effective theory.

Let us start by deriving a form which depends explicitly on the pure spinors $\Psi^\pm$ for the potential ${\cal V}(\Sigma,\calf)$  associated to a  space-time filling D-brane wrapping the generalized cycle     $(\Sigma,\calf)$.
Consider a D-brane wrapping an  $n$-dimensional cycle $\Sigma$ and introduce a complex F-term vector density ${\cal W}_m$, a real D-term density ${\cal D}$ and the scalar density $\Theta$ in the following way
\bea \label{FDformulas}
{\cal W}_m d\sigma^1\wedge\ldots \wedge d\sigma^n &=& \frac{(-)^{n+1}}2P[e^{3A-\Phi}(\imath_m+g_{mk}dy^k\wedge  ) \hat\Psi_2]\wedge e^\calf|_{\rm top}\ ,\cr
{\cal D}d\sigma^1\wedge\ldots \wedge d\sigma^n &=& P[e^{2A-\Phi}{\rm Im}\hat\Psi_1]\wedge e^\calf|_{\rm top}\ ,\cr
\Theta d\sigma^1\wedge\ldots \wedge d\sigma^n &=&  P[e^{4A-\Phi}{\rm Re}\hat\Psi_1]\wedge e^\calf|_{\rm top}\ .
\eea
Note that, if we are not ``too far" from a supersymmetric configuration (which has also an appropriate orientation on $\Sigma$), we can assume that $\Theta> 0$. From the discussion presented in  \cite{lucal}, we can argue that
\bea
\sqrt{\det(P[g]+\calf)}=e^{-4A+\Phi}\sqrt{\Theta^2+e^{4A}{\cal D}^2+2e^{2A}g^{mn}{\cal W}_m\bar{\cal W}_n}\ .
\eea
The complete four-dimensional potential for the D-brane is then given by
\bea                                                                                                                          \label{fullpot}
{\cal V}(\Sigma,\calf)=\int_\Sigma d^n\sigma \sqrt{\Theta^2+e^{4A}{\cal D}^2+2e^{2A}g^{mn}{\cal W}_m\bar{\cal W}_n}-
\int e^{4A}\tilde C\wedge e^\calf\ .
\eea

This potential contains the full nonlinear (static) interactions governing the D-brane.  We want  now to consider the expansion of such a potential around a supersymmetric vacuum configuration $(\Sigma_0,\calf_0)$, which is characterized by the condition ${\cal W}_m(\Sigma_0,\calf_0)={\cal D}(\Sigma_0,\calf_0)=0$. Then, we can consider very small ${\cal W}_m$ and ${\cal D}$ and expand the square root in the potential (\ref{fullpot}). As a result, at the quadratic order in  ${\cal W}^m$ and ${\cal D}$  we obtain the following potential
\bea                                                                                                                          \label{pertpot}
{\cal V}(\Sigma,\calf)&\simeq&\int_\Sigma P[\omega^{(4d)}]\wedge e^\calf+ \frac12\int_\Sigma \frac{d^n\sigma}{\Theta} (e^{4A}{\cal D}^2+2e^{2A}g^{mk}{\cal W}_m\bar{\cal W}_k)=\cr
&=&{\cal V}(\Sigma_0,\calf_0)+ \frac12\int_\Sigma \frac{d^n\sigma}{\Theta} (e^{4A}{\cal D}^2+2e^{2A}g^{mk}{\cal W}_m\bar{\cal W}_k)\ ,
\eea
where in the last step we have used the $d_H$-closedness                                                          of  the generalized calibration $\omega^{\rm (4d)}$ defined in (\ref{4dcal}).

In order to better identify this potential with the standard potential of ${\cal N}=1$ gauge theories, we must introduce  metrics on the spaces\footnote{We indicate with $\Gamma(E)$ the space of  sections of a vector bundle $E$ and with $\Gamma(N,\mathbb{R})$ the space of real functions on a manifold $N$.} $\Gamma(\Sigma,\mathbb{R})$ (that can be identified with the Lie algebra of the four-dimensional gauge group $U(1)^\infty$) and $\Gamma(T_M|_\Sigma)$.
Let us consider first two world-volume functions $f,g\in \Gamma(\Sigma,\mathbb{R})$. We define
\bea                                                                                                                          \label{kmetric}
k(f,g)\equiv \int_\Sigma fg P[e^{-\Phi}{\rm Re}\hat\Psi_1]\wedge e^\calf\ .
\eea
This can be easily seen to be the natural metric for the Lie algebra of the gauge group by expanding the DBI action to write the kinetic term for the four-dimensional field-strength.
Secondly, we introduce the following metric on $\Gamma(T_M|_\Sigma)$
\bea                                                                                                                          \label{mmetric}
G(X,Y)\equiv \int_\Sigma g_{mn}X^m Y^n  P[e^{2A-\Phi}{\rm Re}\hat\Psi_1]\wedge e^\calf\ ,
\eea
that on the other hand defines the natural metric on the space of  four-dimensional scalars (this will become more evident from the following discussions). The above metrics are  non-degenerate for generalized cycles $(\Sigma,\calf)$ not ``too far" from the supersymmetric ones, for which \bea
P[{\rm Re}\hat\Psi_1]\wedge e^\calf|_{\rm top}=\frac{\sqrt{\det(P[g]+\calf)}}{\sqrt{\det P[g]}}d{\rm Vol}_\Sigma\ .
\eea

We can now consider the densities ${\cal D}$ and ${\cal W}_m$ as belonging to the dual of $\Gamma(\Sigma,\mathbb{R})$ and $\Gamma(T_M|_\Sigma)$ by using the natural pairing  given by the ordinary integration (if for example $f\in \Gamma(\Sigma,\mathbb{R})$ and $\theta$ is a dual density, $\langle \theta,f\rangle=\int_\Sigma f\theta$ ). Thus, we can write the potential (\ref{pertpot}) in the form
\bea                                                                                                                          \label{pertpot2}
{\cal V}(\Sigma,\calf)&\simeq& {\cal V}(\Sigma_0,\calf_0)+ \frac12 k^{-1}({\cal D},{\cal D})+G^{-1}(d{\cal W},d\bar{\cal W})\ ,
\eea
where the ${\cal W}_m$'s have been considered as the components of the formal object $d{\cal W}={\cal W}_m dx^m$. As it will be clear from the following sections, we can really think to $d{\cal W}$ as a differential of a proper superpotential ${\cal W}$. Thus, in the expansion of   ${\cal V}$ given in (\ref{pertpot2}) one can recognize a contribution ${\cal V}(\Sigma_0,\calf_0)$, which can be seen as a zero point energy, plus a term that is formally identical to  the standard potential of the  ${\cal N}=1$ theories, given by the sum of the squares of the D- and F-terms.

In order to  identify more explicitly ${\cal W}_m$ and ${\cal D}$ as the actual F- and D-terms of the four-dimensional ${\cal N}=1$ description for space-time filling D-branes,  we can give a look at the form of the supersymmetry transformations of the world-volume fermions. This can be obtained by gauge fixing the $\kappa$-symmetry of the D-brane superactions \cite{ced,bt}. Here we will use the conventions of  \cite{mrvv}, consistently with \cite{lucal}. In particular,  we will use the covariant $\kappa$-fixing explicitly discussed in \cite{mrvv}, where the second Majorana-Weyl spinor is put to zero. Then, the D-brane fermionic degrees of freedom are described by a single ten dimensional Majorana-Weyl spinor $\theta$ and  a background Killing spinors $\varepsilon$ induces a corresponding supersymmetry transformation for $\theta$. When we specialize to the supersymmetry transformations of an ${\cal N}=1$ vacuum configuration, i.e. with constant fields preserving Poincar\'e symmetry, these are given by
\bea                                                                                                                          \label{nonlinsusy}
\delta_\varepsilon\theta&=&\zeta_+\otimes \Big[ \Big(1-\frac{e^{\Phi-4A}\Theta}{\sqrt{\det(P[g]+\calf)}}+i\frac{e^{\Phi-2A}{\cal D}}{\sqrt{\det(P[g]+\calf)}}\Big)\eta^{(1)}_+ +\cr && +\frac{{e^{\Phi-3A}g^{mn}\cal W}_n}{\sqrt{\det(P[g]+\calf)}}\hat\gamma_m\eta_-^{(1)}\Big] +\ {\rm c.c.}\ ,
\eea
where $\zeta_+$ is an arbitrary constant four-dimensional spinor of positive chirality generating the ${\cal N}=1$ supersymmetry (see \cite{lucal} for more details on the notation). If we now consider ${\cal W}_m$ and ${\cal D}$ very small (around a supersymmetric configuration), the above supersymmetry transformation becomes at leading order
\bea                                                                                                                          \label{FDsusy}
\delta_\varepsilon\theta\simeq ie^{-2A}{\cal D}^*(\zeta_+\otimes\eta^{(1)}_+) +e^{-A}{\cal W}^{* n}(\zeta_+\otimes\hat\gamma_m\eta_-^{(1)}) +\ {\rm c.c.}\ ,
\eea
where ${\cal D}^*\in\Gamma(\Sigma,\mathbb{R})$ and ${\cal W}^{*}\in \Gamma(T_M|_\Sigma)$ are the duals associated to ${\cal D}$ and $d{\cal W}$ by using the metrics (\ref{kmetric}) and (\ref{mmetric}) respectively.
In order to give a four-dimensional interpretation of the above supersymmetry transformation, let us first have a look at the bosonic field content living on the D-brane.

By starting from a fixed world-volume field strength configuration $\calf$ on the internal cycle $\Sigma$,  the world-volume  gauge field fluctuations  split in two parts: $a_\mu(x,\sigma)$, with indices along the four flat directions, and $a_\alpha(x,\sigma)$ with indices along the  internal directions wrapped by the cycle. On the other hand, the fluctuations of the brane  can be described by sections  $\phi^{\hat m}(x,\sigma)\in \Gamma(T^\perp_\Sigma)$ of the orthogonal bundle. Note that all the fields depend on $\sigma$, since we are not doing a real dimensional reduction. So, we can think of them as  containing an infinite set of four-dimensional fields. In particular, $a_\mu(x,\sigma)$ gives rise to an infinite set of four-dimensional ${\cal N}=1$ abelian vector multiplets, once completed with the corresponding gaugini that we indicate schematically with $\lambda(x,\sigma)$. On the other hand  $a_\alpha(x,\sigma)$ and $\phi^{\hat m}(x,\sigma)$ should combine to form the bosonic content of an infinite number of chiral multiplets whose  fermionic components we indicate with  $\chi^m$.

We can now  consider more closely the $SU(3)\times SU(3)$ structure associated to the internal manifold $M$. Each $SU(3)$ factor gives a different reduction of the $SO(6)$ structure group of $T_M$, and is an independent symmetry acting separately on $\eta^{(1)}$ and $\eta^{(2)}$ (see e.g. the discussion on \cite{Grana}). Since we are adding to such a background a space-time filling D-brane wrapping an  $n$-dimensional internal cycle, the $SO(1,3)$ symmetry of the four-dimensional part is unbroken, while the two possible $SU(3)$ reductions of the internal  $SO(6)$ structure group are generally spontaneously broken.  Note that the chiral multiplets are the only ones who transform under the $SU(3)$ structure groups of the internal manifold, while the gauge multiplets are singlets with respect to it.

Turning to fermions, before fixing the $\kappa$-symmetry in the superaction of the D-brane, the fermionic content is given by a pair of ten dimensional Majorana-Weyl fermions $\theta^{(1)}$ and $\theta^{(2)}$  of opposite/same chirality in Type IIA/IIB, each transforming under one of the two $SU(3)$'s structure groups. The covariant $\kappa$-fixing used here consists in imposing the condition $\theta^{(2)}=0$, keeping only $\theta\equiv\theta^{(1)}$ as physical degrees of freedom (for which the supersymmetry transformations take the form (\ref{nonlinsusy})). Note that this type of $\kappa$-fixing select the $SU(3)$ associated to $\eta^{(1)}_+$, that we call $SU(3)_1$ from now on,  as the natural one to be used to classify the world-volume fields. Of course, we could have made the alternative (still covariant) choice $\theta^{(1)}=0$,  but this would have been a little less natural since the physical fermion $\theta^{(2)}$ has different chirality for Type IIA and IIB, leading to a somehow less mirror-symmetric description. This  natural selection of  the $SU(3)_1$ structure group will emerge again in the following discussions.

Now, the sixteen components of the ten dimensional fermion $\theta$ splits in the following way under the full structure group $Spin(1,3)\times SU(3)_1$
\bea
{\bf 16}\rightarrow ({\bf 2},{\bf 1})\oplus ({\bf 2},{\bf \bar 3})+{\rm c.c.}\quad .
\eea
The four-dimensional vector fields  transform as singlets under the  internal $SU(3)_1$ structure group, while the scalar fields  transform in the ${\bf 3}+{\bf \bar 3} $ representation.  Then, the fermions in the $({\bf 2},{\bf 1})$ sector must be clearly included in the vector multiples and identified with the gaugini $\lambda$, while the $({\bf 2},{\bf \bar 3})$ sector is given by the fermionic fields $\chi^m$ of the chiral multiplets. Since a base for   the $({\bf 2},{\bf 1})$ sector is given by $\zeta_+\otimes \eta^{(1)}_+$, while a base for the $({\bf 2},{\bf \bar 3})$ sector is given by  $\zeta_+\otimes \hat\gamma_m\eta^{(1)}_- $, we can extract  $\lambda$ and $\chi^m$ from the following splitting of $\theta$
\bea                                                                                                                          \label{fermsplit}
\theta(x,\sigma)=e^{-2A(\sigma)}\lambda(x,\sigma)\otimes\eta^{(1)}_+(\sigma) + \frac{1}{\sqrt2}e^{-A(\sigma)}\chi^m(x,\sigma)\otimes\hat\gamma_m \eta^{(1)}_-(\sigma) +\ {\rm c.c.} \ ,
\eea
where we have indicated explicitly the dependence on the flat and internal world-volume coordinates $x^\mu$ and $\sigma^\alpha$.
The normalizations in (\ref{fermsplit}) have been fixed by requiring that $\lambda$ and $\chi^m$ must have canonical kinetic term, using the internal metrics (\ref{kmetric}) and (\ref{mmetric}) for $\lambda$ and $\chi^m$ respectively. Indeed, the explicit form of the quadratic fermionic terms on a general background was found in \cite{mms}  and using this it is easy to see that the kinetic term for $\lambda$ and $\chi^m$ are given by
\bea
{\cal L}^F_{\rm kin}&=&i\int_\Sigma \bar\lambda\gamma^\mu\partial_\mu\lambda P[e^{-\Phi}{\rm Re}\hat\Psi_1]\wedge e^\calf + i\int_\Sigma g_{mn}\bar\chi^m\gamma^\mu\partial_\mu\chi^n P[e^{2A-\Phi}{\rm Re}\hat\Psi_1]\wedge e^\calf =\cr
&=& ik(\bar\lambda,\gamma^\mu\partial_\mu\lambda)+iG(\bar\chi,\gamma^\mu\partial_\mu\chi)\ .\eea
Thus, from (\ref{FDsusy}) we obtain the standard supersymmetry transformations (for constant field configurations)
\bea
\delta_{\zeta}\lambda= i{\cal D}^*\zeta\quad,\quad \delta_{\zeta}\chi^m= \sqrt2\,{\cal W}^{*m}\zeta\ .
\eea
Even if we have not computed precisely the dimensional reduction and appropriately organized all the tower of KK fields in supersymmetric multiplets, we can nevertheless conclude that from the four-dimensional point of view one can indeed consider ${\cal D}$ as a D-term and ${\cal W}_m$ as an F-term, motivating the names used to label the supersymmetry conditions (\ref{FDterms}). 

Let us stress again that the above analysis was done in the `linearized' approximation where ${\cal D}$ and ${\cal W}_m$ are small and we expect the theory to be well described by a standard ${\cal N}=1$ theory. Note that for the full DBI theory the vanishing of the D-term ${\cal D}$ alone is not enough to insure the vanishing of the gaugini supersymmetry transformations (see (\ref{nonlinsusy})). On the other hand, the vanishing of the F-term ${\cal W}_m$ alone {\em is} enough to insure that vanishing of supersymmetry transformation of the fermions in the chiral multiplets. We will see in the next section that the F-terms do indeed have a clear non-linear validity, by constructing the explicit complete superpotential generating them. A nonlinear interpretation of the D-term will arise in section \ref{symplectic}.

\section{Superpotential for D-branes on ${\cal N}=1$ vacua}
\label{supsec}

In section \ref{sec1} we have recalled  how supersymmetric D-branes in D-calibrated ${\cal N}=1$   backgrounds  can be seen as calibrated D-branes with respect to properly  defined generalized calibrations. This condition is in turn equivalent (up to an appropriate orientation choice) to the pair of conditions given in (\ref{FDterms}) that, as discussed in section \ref{sec2}, may be seen as the F-flatness and D-flatness conditions in the language of the ${\cal N}=1$ four-dimensional formulation. In this section we show how the F-flatness condition in (\ref{FDterms}) can be further extracted from a corresponding superpotential. This provides a generalization and, in some sense, a reformulation in a unified language, of previous superpotentials obtained in the Calabi-Yau case along the lines of the two-cycle case considered in \cite{wittenQCD} (see for example \cite{thomas} for a general discussion in the Calabi-Yau case). Our superpotential is obviously  applicable also in the limiting case of ${\cal N}=2$ backgrounds with only nontrivial NS fields and their simplest subcase in which the internal manifold reduces to a standard Calabi-Yau.

Let us start by discussing the space of relevant degrees of freedom. We take as configuration space ${\cal C}$ the space of all the generalized cycles $(\Sigma,\calf)$ quotiented by the group of internal world-volume diffeomorphisms $Diff(\Sigma)$. The space ${\cal C}$ can be properly identified with the space of the scalar fields in the four-dimensional description of the system. The world-volume gauge transformations that depends only on the $\Sigma$-coordinates (and not on the ones filling the four flat directions) corresponds to an infinite family of abelian {\em rigid} symmetries of the scalar field space ${\cal C}$, that are gauged in the full theory. The tangent space $T_{\cal C}$ to ${\cal C}$ should describe the  infinitesimal deformations of the embedded submanifolds  and of the world-volume field strength on them.

We first describe the deformations of the field strength $\calf$ due to the deformations of the world-volume  gauge field, while keeping the embedded submanifold $\Sigma$ fixed. Since $\calf$ must satisfy the generalized Bianchi identity $d\calf=P_\Sigma[H]$, an infinitesimal variation of $\calf$ must be of the form $\delta \calf=da$, where $a\in \Gamma(T^*_\Sigma)$ is a globally defined one-form on $\Sigma$. As we have said, the infinitesimal  gauge transformations $a\rightarrow a+d\lambda$, with $\lambda\in \Gamma(\Sigma,\mathbb{R})$, can be considered as the rigid transformations in the four-dimensional description of the system, that are gauged by the coupling to the four-dimensional vector fields. Secondly, we consider the general class of deformations of the submanifold $\Sigma$ in $M$ generated by a section $X\in\Gamma(T_M|_\Sigma)$ of the bundle $T_M$ restricted to $\Sigma$. Note that such a a deformation induces also a corresponding infinitesimal transformation $\delta\calf=P_\Sigma[\imath_X H]$ on the world-volume field-strength. A more detailed  discussion on the infinitesimal deformations of the generalized cycle $(\Sigma,\calf)$ is contained in  appendix \ref{appA}.

Now, not all these infinitesimal deformations are physically distinguishable since some could be related by an infinitesimal $\Sigma$-diffeomorphism. At the infinitesimal level, a $\Sigma$-diffeomorphism can be identified by a vector field $v\in \Gamma(T_\Sigma)$. Then, by associating $v$ to its push-forward in $\Gamma(T_M|_\Sigma)$, we obtain that the infinitesimal transformations of the form
\bea
X= v\quad,\quad \delta \calf =
d\imath_{v}\calf+P[\imath_{v}H]\ ,
\eea
must be considered as non-physical, and must be quotiented out.
Then we must consider the following  ``gauge" equivalence between two infinitesimal deformations of the generalized cycle $(\Sigma,\calf)$
\bea                                                                                                                          \label{gentangauge}
X\simeq X+v\quad,\quad a\simeq a +\imath_{v}\calf\ .
\eea
Such  identifications can be appropriately described in generalized geometry terms, by recalling the definition of generalized tangent bundle $\tau_{(\Sigma,\calf)}$ of a generalized cycle $(\Sigma,\calf)$ given in \cite{gualtieri}:
\bea                                                                                                                          \label{gentang}
\tau_{(\Sigma,\calf)}=\{v+\eta\in  T_\Sigma\oplus T_M^*|_\Sigma \ :\ P_\Sigma[\eta]=\imath_v\calf \}\ .
\eea
 From (\ref{gentang}) it is clear that the  tangent space $T_{\cal C}$ of ${\cal C}$ at a ``point" $(\Sigma,\calf)$ can be
 identified with the space of sections of the vector bundle ${\cal N}_{(\Sigma,\calf)}\equiv
 (T_M\oplus T_M^*)|_\Sigma/\tau_{(\Sigma,\calf)}$, that we call the {\em generalized normal  bundle} of $(\Sigma,\calf)$.

We are now ready to rewrite the F-flatness condition in (\ref{FDterms}) in a form that can  be more immediately recognized as coming from a superpotential.  In order to do this, let us start by splitting the F-flatness condition in two, by projecting it in the orthogonal and tangent directions  using the metric structure of the background. We consider first an arbitrary  vector field $X_{\perp}\in T^\perp_\Sigma$ orthogonal to $\Sigma$. If we consider the projection of the F-flatness condition in (\ref{FDterms}) along $X_\perp$, we obtain
\bea                                                                                                                          \label{norm}
P_{\Sigma}[\imath_{X_\perp}\hat\Psi_2]\wedge e^\calf |_{\rm top}=0\ .
\eea
Secondly,  we consider an arbitrary section $X_{||}$ of the tangent bundle $T_{\Sigma}$ of $\Sigma$. Projecting the F-flatness condition along $X_{||}$ gives the equation
\bea                                                                                                                          \label{tang}
P[X_{||}^*\wedge \hat\Psi_2 + \imath_{X_{||}}\hat\Psi_2]\wedge e^\calf|_{\rm top}=0\ ,
\eea
where $ X_{||}^*$ is the one-form canonically associated to $X_{||}$ through the world-volume metric $P[g]$. It is easy to see that it is possible  to write this equation in the following equivalent way (we use internal world-volume coordinates $\sigma^\alpha$)
\bea                                                                                                                          \label{intang}
(P[g]+\calf)_{\alpha\beta}X_{||}^\beta d\sigma^\alpha \wedge P[\hat\Psi_2]\wedge e^{\calf}=0\ .
\eea
Thus, since $P[g]+\calf$ is non-degenerate for non-degenerate brane configurations, we can rewrite the F-flatness condition in (\ref{FDterms}) as the following pair of conditions
\bea                                                                                                                          \label{geomF}
&&P_\Sigma[\hat\Psi_2]\wedge e^{\calf}|_{{\rm top}-1}=0\ ,\cr
 &&P_{\Sigma}[\imath_{X}\hat\Psi_2]\wedge e^\calf |_{\rm top}=0\ ,
 \eea
where now $X$ is an arbitrary section of $T_M|_\Sigma$. Note that, by using the first  of the F-flatness conditions (\ref{geomF}), the second can in fact be though as $X$ was actually a section  the canonical normal bundle  ${\cal N}_\Sigma=T_M/T_\Sigma$, since it is left invariant if we substitute $X$ with $X+v$ for any $v\in\Gamma(T_\Sigma)$.

We can now present the superpotential generating the F-flatness conditions (\ref{geomF}), postponing to section \ref{hol} the discussion of how it can be actually considered as holomorphic. We want to define a superpotential ${\cal W}$ as functional of the pair $(\Sigma,\calf)$ defining the internal configuration of the four-dimensional space-time filling D-brane. In order to define such a functional, we need to introduce a fixed reference generalized cycle $(\Sigma_0,\calf_0)$ which is smoothly related to  $(\Sigma,\calf)$. More precisely, we require that $(\Sigma_0,\calf_0)$ is in the same {\em generalized homology class}  of $(\Sigma,\calf)$, that is there must exist a chain ${\cal B}$ and a field strength $\tilde\calf$ on it (satisfying $d\tilde\calf=P_{\cal B}[H]$) such that\footnote{In the case in which $\Sigma$ has zero homology class we can take an empty $\Sigma_0$ and the conditions (\ref{bouncond}) can be simplified to the pair of conditions $\partial{\cal B}=\Sigma$ and $P_\Sigma[\tilde\calf]=\calf$.}
\bea                                                                                                                          \label{bouncond}
\partial {\cal B}=\Sigma-\Sigma_0\quad,\quad P_\Sigma[\tilde\calf]=\calf\quad {\rm and}\quad P_{\Sigma_0}[\tilde\calf]=\calf_0\ .
\eea
Then the superpotential whose critical points are given by the F-flatness conditions  (\ref{geomF}) can  be defined by
\bea                                                                                                                          \label{suppot}
{\cal W}(\Sigma,\calf)-{\cal W}(\Sigma_0,\calf_0)=\frac12\int_{\cal B}P[e^{3A-\Phi}\hat\Psi_2]\wedge e^{\tilde\calf} \ .
\eea
The formula (\ref{suppot}) defines the superpotential ${\cal W}(\Sigma,\calf)$ up to an additive constant, whose indeterminacy comes from the arbitrary choice of $(\Sigma_0,\calf_0)$ and also by the possible nontrivial topology of the background\footnote{If for example the homology group $H_{{\rm dim}\Sigma+1}(M,\mathbb{R})$ is non-zero, there are possible non-homologous choices of ${\cal B}$ (for fixed boundary conditions). The choice of a different class in $H_{{\rm dim}\Sigma+1}(M,\mathbb{R})$ gives a shift of ${\cal W}$ by a constant.}. We will see in the next section how we can give to  ${\cal B}$ and $\tilde\calf$ defined in (\ref{bouncond}) a clear physical interpretation.

It is immediate to obtain (\ref{geomF}) as critical point conditions for the superpotential (\ref{suppot}). Indeed, consider any generalized normal vector $[(X,a)]\in \Gamma({\cal N}_{(\Sigma,\calf)})$, associated to the representative $(X,a)$. Then, the infinitesimal variation of ${\cal W}$ defined by $[(X,a)]$ is given by
\bea                                                                                                                          \label{infvarW}
\delta{\cal W}=\frac12\int_\Sigma\Big\{ P[e^{3A-\Phi}\imath_X\hat\Psi_2] +a \wedge  P[e^{3A-\Phi}\hat\Psi_2]   \Big\}\wedge e^\calf\ .
\eea
Note that clearly the above infinitesimal variation is invariant under the substitution $(X,a)\rightarrow (X+v,a+\imath_v\calf)$, for any $v\in\Gamma(T_\Sigma)$, and thus it is well defined for the equivalence class $[(X,a)]$. From (\ref{infvarW}) it is clear that the superpotential critical points are defined by the conditions (\ref{geomF}). Note also that the two terms  (\ref{infvarW}) can be directly identified with the left hand side of (\ref{norm}) and (\ref{intang}) by choosing a gauge with $X=X_\perp$ orthogonal to $\Sigma$ and making the identification
\bea                                                                                                                          \label{tangrel}
a=(P[g]+\calf)\cdot X_{||}\ .
\eea
This provides an explicit identification of $T_{\cal C}|_{(\Sigma,\calf)}=\Gamma({\cal N}_{(\Sigma,\calf)})$ with $\Gamma(T_M|_\Sigma)$, which uses in an essential way the background metric. In the following we will often use this identification, which will allow us to introduce an almost complex and a symplectic structure on ${\cal C}$.

Note that, as the F-flatness conditions  in the form (\ref{geomF}), the superpotential does not depend on the full $SU(3)\times SU(3)$ structure (which contains also the metric structure) characterizing the internal manifold of the ${\cal N}=1$ backgrounds we are considering, but it involves only the integrable pure spinor. This result  could be seen as a generalization of the ``decoupling statement" presented in \cite{quintic}, which asserts that the superpotentials governing D-branes in Calabi-Yau spaces depend only on the background complex structure and not on the K\"ahler structure for B-branes, and vice-versa for A-branes. 
The same superpotential may be used to describe also topological D-branes \cite{Kapustin,zabzine,Zucchini,Li,Kapustin:2005vs} for the underlying topological model \cite{Kapustin,kapu2,gmpt2}, since its form is clearly valid for any generalized Calabi-Yau structure, as defined by Hitchin in \cite{hitchin}.  Namely, for any generalized Calabi-Yau manifold  defined by a $d_H$-closed pure spinor $\psi$, we can introduce a variational problem to characterize the generalized complex submanifolds $(\Sigma,\calf)$ as the extrema of the functional
\bea                                                                                                                          \label{topsuppot}
F(\Sigma,\calf)=\frac12\int_{{\cal B}}P[\psi]\wedge e^{\tilde{\calf}}\ ,
\eea
where ${\cal B}$ and $\tilde\calf$ are defined as for the specific case of the ${\cal N}=1$ backgrounds considered before.

The above superpotentials can be written directly in terms of the generalized cycle $(\Sigma,\calf)$ by using the $d_H$-closedness                                                          of  $e^{3A-\Phi}\hat\Psi_2$ (or analogously of $\psi$ in (\ref{topsuppot})). Indeed, we can locally write  $e^{3A-\Phi}\hat\Psi_2=d_H\chi$, where $\chi$ is again a polyform, and then
\bea                                                                                                                          \label{localsuppot}
{\cal W}(\Sigma,\calf)=\frac12\int_\Sigma P[\chi]\wedge e^\calf +\ {constant}\ .
\eea
Note that the  expression (\ref{suppot}) for the superpotential   is completely analogous to the CS term of the D-brane action and like that it  is meaningful even if the `potential' polyform $\chi$ is not generally globally defined.

To close this section, let us stress that till now we have  deliberately ignored the tension $\mu_p=2\pi(2\pi\sqrt{\alpha^\prime})^{-(p+1)}g_s^{-1}$ of the D$p$-brane we are considering (i.e., we have fixed $\mu_p=1$). The tension should be of course reintroduced to have the correct dependence on the fundamental quantities $\alpha^\prime$ and $g_s$. 
The canonically normalized superpotential ${\cal W}_{\rm can}$ which includes the correct dependence on the tension is given by
\bea                                                                                                                          \label{cansuppot}
{\cal W}_{\rm can}=\mu_p{\cal W}\ ,
\eea
as follows directly from the form of the potential (\ref{pertpot2}), since the canonically normalized potential and metric are given by ${\cal V}_{\rm can}=\mu_p{\cal V}$ and  $G_{\rm can}=\mu_p G$ respectively, where $G$ is defined in (\ref{mmetric}).
As we will see in the following section, the superpotential (\ref{suppot}) can be derived from an argument involving domain walls, which also gives an alternative consistency check of the above normalization of the superpotential. In the following we will re-intruduce the correct dependence on the tension only when needed, continuing to neglect it in  most of the discussions.

\section{Superpotential from domain walls}
\label{supDW}

In the previous section we have shown how to obtain the F-flatness conditions in (\ref{FDterms}) or equivalently (\ref{geomF}) as the conditions defining the critical points of the superpotential (\ref{suppot}). In this section we  use a physical argument that leads directly to the above superpotential, confirming its validity from a more physical point of view. This can be seen as a generalization to the D-brane context of the standard Gukov-Vafa-Witten argument used to derive the superpotential  governing supergravity compactifications with fluxes \cite{gukov1,gukov2}. In particular we will see how the domain wall generalized calibration given in (\ref{others}), being naturally related to the integrable generalized complex structure of the background, is also naturally related to the F-term associated to the space-time filling D-brane. Along  the way, it will allow  to check  the canonical normalization of  the superpotential given in (\ref{cansuppot}).

For a given space-time filling D-brane, consider two supersymmetric configurations $(\Sigma_1,\calf_1)$ and $(\Sigma_2,\calf_2)$  that belong to the same generalized homology class.  These can be seen as ${\cal N}=1$ vacua of the effective ${\cal N}=1$  four-dimensional supersymmetric theory governing the D-brane dynamics. Then, on general grounds, we expect that a domain wall  interpolating between the two vacua can exist. Such a domain wall configuration can be constructed in the following way. Take a  D-brane filling the half of space-time with positive third space coordinate, $x^3> 0$,  and wrapping the supersymmetric generalized cycle $(\Sigma_1,\calf_1)$,  and another D-brane (of the same kind)  filling the other half of space-time with  $x^3< 0$ and  wrapping the other supersymmetric generalized cycle $(\Sigma_2,\calf_2)$. These two D-brane configurations with boundary  $\mathbb{R}^{1,2}\times \{x^3=0\}$ can be glued together in a consistent way by filling the common boundary with another D-brane (again, of the same kind) wrapping a generalized cycle $({\cal B},\tilde\calf)$ defined by a chain ${\cal B}$  with boundary such that $\partial{\cal B}=\Sigma_1-\Sigma_2$ and a world-volume field-strength $\tilde\calf$ such that $P_{\Sigma_1}[\tilde\calf]=\calf_1$ and   $P_{\Sigma_2}[\tilde\calf]=\calf_2$.   The choice of the field-strength $\tilde\calf$ is the right one to glue together the three D-brane  configurations with boundaries in such a way that the usual anomaly terms coming from the boundaries of each D-brane  \cite{strom,town,cosmic} cancel each other. In order to see it, let us write the complete set of Ramond-Ramond potentials in the form $C=\sum_{k} C_{(k)}$, where $k$ is odd in Type IIA and even in Type IIB, and consider the general gauge transformation $\delta C=d_H\lambda$, where $\lambda=\sum_k \lambda_{(k-1)}$. The CS term in the action of the two half space-time filling D-branes transforms in the following way
\bea                                                                                                                          \label{anomaly}
\delta S_{1}^{\rm CS}+\delta S_2^{\rm CS}&=&\delta \int_{\mathbb{R}^{1,2}\times \mathbb{R}_+\times \Sigma_1} P[C]\wedge
e^{\calf_1} +\delta \int_{\mathbb{R}^{1,2}\times \mathbb{R}_-\times \Sigma_2} P[C]\wedge
e^{\calf_2}=\cr
&=& -\int_{\mathbb{R}^{1,2}\times \Sigma_1} P[\lambda]\wedge
e^{\calf_1} +\int_{\mathbb{R}^{1,2}\times \Sigma_2} P[\lambda]\wedge
e^{\calf_2}\ .
\eea
Then the  gauge symmetry is broken by the boundary terms if we consider the two half space-time filling D-branes alone. However, the introduction of the domain wall D-brane located at $x^3=0$ as described above provides the necessary counterterm  to reabsorb   the undesired terms in (\ref{anomaly}). Indeed, the domain wall D-brane action  contains the CS term
\bea
S^{\rm CS}_{\rm DW}=\int_{\mathbb{R}^{1,2}\times {\cal B}}P[C]\wedge e^{\tilde\calf}\ ,
\eea
and it is easy to see that its variation under the gauge transformation $d_H\lambda$ exactly cancels the two terms in (\ref{anomaly}).

Now, from general arguments in ${\cal N}=1$ supersymmetric field theories (see e.g. \cite{cvetic}), it is known that the tension of a BPS  domain wall  is simply given  by
\bea                                                                                                                          \label{FTDW}
T_{\rm DW}=2{\rm Re}(e^{i\theta}\Delta {\cal W})\ ,
\eea
where $\Delta{\cal W}={\cal W}_1-{\cal W}_2$ is the superpotential difference of the two different vacua and $\theta$ define a constant phase related to the preserved half of supersymmetry. On the other hand, from our D-brane construction the field theory domain wall tension should be  exactly given by the effective tension of a supersymmetric configuration for the D-brane domain wall introduced above.  But, from the general discussion of \cite{lucal} reviewed in section \ref{sec1}, we know that such a supersymmetric domain wall D-brane must wrap a generalized cycle   calibrated with respect to the generalized calibration $\omega^{\rm (DW)}$ written in (\ref{others}). From this, we immediately obtain that the tension of the BPS D-brane domain wall is given by
\bea                                                                                                                          \label{DDWtension}
T_{\rm DW}=\int_{\cal B}P[e^{3A-\Phi}{\rm Re}(e^{i\theta}\hat\Psi_2)]\wedge e^{\tilde\calf}\ ,
\eea
where again $\theta$ defines the preserved supersymmetry.
Comparing this expression with the one given in (\ref{FTDW}), one can immediately extract the the form of the superpotential as written in (\ref{suppot}) (again defined up to an additive constant). Furthermore, by reintroducing the neglected tension $\mu_p$ in front of the right hand side of (\ref{DDWtension}), we obtain the canonically normalized superpotential (\ref{cansuppot}). Note that, from the general analysis of \cite{lucal},  the fact that the domain wall D-brane is calibrated with respect to the generalized calibration $\omega^{\rm (DW)}$ of (\ref{others}) implies also that $P_{\cal B}[e^{3A-\Phi}{\rm Im}(e^{i\theta}\hat\Psi_2)]\wedge e^{\tilde\calf}|_{\rm top}=0$. Thus, as in field theory, the phase $\theta$ in (\ref{DDWtension}) is directly related to the phase of superpotential difference, i.e. $e^{-i\theta}=\Delta {\cal W}/|\Delta{\cal W}|$, so that
\bea
T_{\rm DW}=2|\Delta{\cal W}|=|\int_{\cal B}P[e^{3A-\Phi}\hat\Psi_2]\wedge e^{\tilde\calf}|\ .
\eea

\section{Fayet-Iliopoulos terms and cosmic strings}\label{Dstrings}

In the previous section we have seen how the well known relation between the superpotential of an $\caln=1$ theory and supersymmetric domain walls can be exactly reproduced in our D-brane context by using the calibration $\omega^{\rm (DW)}$ defined in (\ref{others}) for D-branes filling only three flat space-time directions.

It this section we will discuss how, on the other hand, our D-terms are related to the  another possible solitonic objects allowed by an ${\cal N}=1$ theory, namely  cosmic strings\footnote{Using this name, we implicitly  refer to cosmological scenarios obtained from flux compactifications. In the context of the gauge/gravity correspondence, these effective string configurations  can be also seen as proper solitonic objects of rigid supersymmetric theories.}. There has been a lot of recent activity focused on the embedding of these  kind of solitons into string theory (for a review see for example \cite{polchistrings}). In particular, in \cite{toinestring} it has been stressed how the only allowed supersymmetric cosmic string solutions  of four-dimensional  ${\cal N}=1$ supergravity must have a vanishing F-term and can exist thanks to D-terms with a non-vanishing constant Fayet-Iliopoulos (FI) term. Furthermore the authors of \cite{toinestring} proposed an identification of the $\caln=1$ four-dimensional supergravity they started from with  the  effective supergravity  theory describing some main features of a  D$\bar{\rm D}$-brane pair  filling the four flat space-time dimensions and wrapping some internal cycle  (see also the related discussions in \cite{toine2,dvali}) . Our formalism allows to give a non-trivial explicit argument in favor of this proposal and a direct D-brane derivation of some of the results of \cite{toinestring} (see also \cite{lawrence,halyo}).

Let us start by considering a single space-time filling D$p$-brane wrapping an internal $n$-dimensional generalized cycle $(\Sigma,\calf)$. The crucial observation is that the D-flatness condition $\cald(\Sigma,\calf)=0$ [the D-term $\cald$ is defined in (\ref{FDformulas})] can be satisfied only if $\int_\Sigma \cald d^n\sigma=0$. By recognizing in $\cald$ the presence of the string generalized calibration $\omega^{\rm(string)}$ written in (\ref{others}), which is $d_H$-closed, we immediately see  that this condition is topological, i.e. does not change if we continuously deform $(\Sigma,\calf)$. Then, from the analysis of section \ref{sec2}, it is natural to identify the constant (reintroducing the tension of the D-brane)
\bea\label{FI}
\xi\equiv \mu_p\int_\Sigma \cald d^n\sigma
\eea
with the FI term of the lowest Kaluza-Klein  four-dimensional $U(1)$ gauge field. Indeed, the corresponding gauge group has no associated charged  chiral fields   and thus the necessary requirement for having a supersymmetric vacuum is that $\xi=0$. Note that even  if $\cald$ was identified as a D-term expanding the action around a supersymmetric configuration, the fact that $\xi$ defined in (\ref{FI}) is constant for {\em any} configuration supports the idea that its identification with an effective  FI term should indeed be more general. This will be confirmed by the following analysis.

Take a space-time filling D$p$-brane wrapping a generalized cycle $(\Sigma,\calf)$ such that $\xi\neq 0$. As we have said, this cannot admit a supersymmetric configuration (at least considering only classical deformations). However, we can add an anti $\bar{\rm D}p$-brane wrapping the same internal generalized cycle $(\Sigma,\calf)$. As a consequence,  the resulting spectrum on the branes includes  now also a complex tachyon which is charged  under the combination $A^{(1)}-A^{(2)}$ of the two gauge fields   $A^{(1)}$ and $A^{(2)}$ living on  the two branes. Thus, from the discussion of the previous paragraph, it seems reasonable to conclude that the lowest Kaluza-Klein mode of the diagonal $U(1)$ gauge group under which the tachyon is charged has $\xi$ as  non-vanishing FI term. The (unstable) system then admits  a vortex solution  \cite{sen} that can be identified with a D$(p-2)$--brane filling only two flat space-time directions and wrapping the internal $(\Sigma,\calf)$-cycle, thus leaving an effective cosmic string.
From the analysis of \cite{lucal}, we can immediately conclude that the resulting cosmic string is supersymmetric if and only if it is calibrated with respect to the generalized calibration $\omega^{\rm(string)}$. This implies that the cosmic sting tension is given by
\bea\label{cstringt}
T_{\rm string}=\mu_{p-2}\int_\Sigma \omega^{\rm (string)}\wedge e ^\calf=(2\pi)^2\alpha^\prime\xi\ .
\eea

On the other hand, since we are considering $\caln=1$ backgrounds, the D$\bar{\rm D}$-brane system  should be described by a four-dimensional $\caln=1$ low energy effective theory. Moreover, since we consider BPS cosmic strings, their tension computed in (\ref{cstringt}) using a probe D$(p-2)$--brane  should be reproduced by the four-dimensional  results of \cite{toinestring}.  Indeed, to recognize the perfect agreement it is enough to remember that in the description given in section \ref{sec2} we have used fluctuating fields with the dimension of a length. The standard dimensions for the fields are obtained by simply rescaling them by $2\pi\alpha^\prime$.  This induces a corresponding rescaling $\xi\rightarrow \xi/2\pi\alpha^\prime$ of the FI term. Thus, in terms of the proper dimensional FI term, the cosmic string tension reads $T_{\rm string}=2\pi\xi$, which is exactly reproduced by the effective supergravity calculation \cite{bbs,jose,toinestring}.

Our argument also allows one to obtain from a purely D-brane setting the observation of \cite{toinestring} that for BPS cosmic strings of an ${\cal N}=1$  four-dimensional supergravity the F-term must vanish identically. Indeed, from the  discussion of \cite{lucal} it follows  that the calibration condition on the generalized cycle $(\Sigma,\calf)$ wrapped by the D-brane forming a BPS cosmic string implies also that $(\Sigma,\calf)$ must be a generalized complex submanifold, i.e. the F-term must vanish identically so that the superpotential (\ref{suppot}) is extremized everywhere.

Let us stress another outcome of our approach. The system constituted by a D$\bar{\rm D}$-brane pair  added to an  $\caln=1$ background should be described by an effective $\caln=1$ supergravity theory like the one considered in \cite{toinestring}.
As  is clear from the above analysis, we can obtain an effective cosmic string as a tachyonic vortex  on a D$\bar {\rm D}$-brane pair  only if these space-time filling branes wrap an internal generalized cycle $(\Sigma,\calf)$ that {\em cannot} be deformed in such a way that the two D-branes, taken singularly,  become supersymmetric. In few words, we must start from a pair of  non-supersymmetric space-time filling  D-branes if we want to create a cosmic string from tachyon condensation. Vice-versa, if we start from  supersymmetric D$p$-branes, then tachyon condensation cannot give rise to any supersymmetric   D$(p-2)$--brane  configuration wrapping a generalized cycle homologous to $(\Sigma,\calf)$. This is an immediate consequence of the fact  that in general parallel  D$p$- and D$(p-2)$--branes do not separately preserve any common  supersymmetry (but generally form a proper bound state). In our case, the ${\cal N}=1$ supersymmetry of the background implies that a  generalized $n$-cycle cannot be contemporary homologous to generalized calibrated cycles for both D$(3+n)$- and D$(1+n)$-branes.

This last  conclusion cannot be  extended to the particular subcases where the RR fields are switched off and  the background preserves $\caln=2$ supersymmetry. Indeed, in these cases we have an arbitrary phase entering the generalized calibrations  (that can be adjusted giving a different preserved internal supersymmetry) and the condition for a generalized cycle to be calibrated does not depend on the number of filled flat directions \cite{mmms,koerber}. However, a supersymmetric D$(1+n)$-brane wrapping a generalized  $n$-cycle preserves exactly the $\caln=1$ supersymmetry that is broken by a D$(3+n)$-brane wrapping the same generalized  $n$-cycle. The associated non-linearly realized supersymmetry on the world-volume of the D$(3+n)$-branes constituting the D$\bar{\rm D}$-brane pair should then be associated to a FI term $\xi$ in a four-dimensional $\caln=1$ description of the system, as happens for $\caln=1$ backgrounds. Then, the above analysis for  $\caln=1$ backgrounds can be repeated with no changes giving again $T_{\rm string}=2\pi\xi$. It would be interesting to understand better the relation between the D-brane picture and a complete ${\cal N}=2$ four-dimensional  supergravity description of one-half BPS cosmic strings, like for example the one presented in \cite{alessio}.


\section{Holomorphicity of the superpotential}
\label{hol}

We can now pass to the discussion of the holomorphic structure of the superpotential introduced in section \ref{supsec}. More precisely, we will  introduce an almost complex structure  on the space ${\cal C}$ of the generalized cycles $(\Sigma,\calf)$ with respect to which the superpotential is holomorphic, i.e. it is annihilated by the $(0,1)$ vectors on ${\cal C}$. Since the space of possible deformations is infinite dimensional, we will work quite at the formal level treating it as finite dimensional, neglecting possible related subtleties.  Furthermore, we shall not worry about the integrability of the almost complex structures introduced. Such an issue is already present for example in the study of Lagrangian
submanifolds \cite{thomas}, but is not so crucial for the following discussion.

Let us start by recalling that the internal manifold $M$ has an integrable generalized complex structure ${\cal J}_2$ associated to the integrable pure spinor $\Psi_2$. It is clearly not sufficient by itself to induce an almost complex structure (integrable or not) on  ${\cal C}$. However,   it {\em does} define a natural almost complex structure, in the sense of an endomorphism of the tangent bundle that squares to minus one, if we restrict $T_{\cal C}$ to the subspace ${\cal C}_{\rm hol}\subset {\cal C}$  of the generalized complex submanifolds. As we have seen in section \ref{sec2}, ${\cal C}_{\rm hol}$ can be characterized as the space of  critical points of the superpotential (\ref{suppot}). Indeed, by definition a generalized cycle $(\Sigma,\calf)$ is complex if the associated tangent bundle $\tau_{(\Sigma,\calf)}$ is stable under ${\cal J}_2$. As a consequence, ${\cal J}_2$ defines a natural almost complex structure on the generalized normal bundle  ${\cal N}_{(\Sigma,\calf)}$ and then on the subset ${\cal C}_{\rm hol}$ of ${\cal C}$  using the identification $T_{\cal C}|_{(\Sigma,\calf)}=\Gamma({\cal N}_{(\Sigma,\calf)})$.

 Now, we would like to introduce an appropriate  (almost) complex structure $\mathbb{J}$ on $T_{\cal C}$ that should provide an  extension to the whole ${\cal C}$ of the complex structure properly defined only on  ${\cal C}_{\rm hol}$. Furthermore, the metric introduced in (\ref{mmetric}) will turn out to define an associated  Hermitian metric on ${\cal C}$.
 In order to do it we must first of all use the  generalized metric structure \cite{gualtieri} on $T_M\oplus T_M^*$, that in our case  is ultimately given by the metric $g$ of $M$, to find a good coordinatization of ${\cal N}_{(\Sigma,\calf)}$.  Using the metric $g$ we can split $T_M|_\Sigma$ in the sum of the tangent and orthogonal bundles to $\Sigma$, $T_M|_\Sigma=T_\Sigma\oplus T^\perp_\Sigma$. Then, we can give a global of  splitting of $T_M\oplus T_M^*|_\Sigma$ appearing in the short exact sequence
 \bea
0\rightarrow \tau_{(\Sigma,\calf)}\rightarrow T_M\oplus T_M^*|_\Sigma\rightarrow {\cal N}_{(\Sigma,\calf)}\rightarrow 0\ ,
 \eea
by using the  $\Sigma$-diffeomorphism invariance to select $X=X_{\perp}\in T^\perp_\Sigma$ in the equivalence class $[(X,a)]\in {\cal N}_{(\Sigma,\calf)}$. This allows to identify ${\cal N}_{(\Sigma,\calf)}$ with $T^\perp_\Sigma \oplus T^*_\Sigma$ and then a general tangent vector of $T_{\cal C}|_{(\Sigma,\calf)}$ can be identified by a pair $(X_\perp,a)\in \Gamma(T^\perp_\Sigma \oplus T^*_\Sigma )$. We can also see this vector as a vector field $X= X_{||}+X_{\perp}\in \Gamma(T_M|_\Sigma)$, using the identification
$a=(P[g]+\calf)\cdot X_{||}$ already introduced in (\ref{tangrel}) to relate the variation of the superpotential to the form of the F-flatness written in (\ref{FDterms}).  At this point we must recall that for our ${\cal N}=1$ backgrounds one can use the  internal spinors $\eta_+^{(1)}$ and  $\eta_+^{(2)}$  to construct a pair of almost complex structures $(J_1)_m{}^n=-(i/|a|^2)\eta_+^{(1)\dagger} \hat\gamma_{m}{}^n\eta_+^{(1)}$ and $(J_2)_m{}^n=-(i/|a|^2)\eta_+^{(2)\dagger} \hat\gamma_{m}{}^n\eta_+^{(2)}$ on $M$  (see \cite{gmpt,gmpt2}, and \cite{lucal} for the conventions used here). Moreover the internal metric $g_{mn}$ is Hermitian with respect to both of them. These almost complex structures define also the null spaces of the two pure spinors $\Psi^\pm$ since
\bea                                                                                                                          \label{null}
(1+iJ_1)_m{}^n(\imath_n+g_{nk}dy^k\wedge)\Psi^{\pm}=0\quad,\quad (1\mp iJ_2)_m{}^n(\imath_n-g_{nk}dy^k\wedge)\Psi^{\pm}=0\ .
\eea
Note that $J_1$ is somehow  selected by the property that its $+i$ eigenspace defines through the above equations the (complex) three dimensional space given by the intersection of the two null subspaces of the two pure spinors $\Psi^\pm$. Indeed, by looking at the F-flatness conditions as written  in (\ref{FDterms}), it is clear that $J_1$ plays a particular role.
We are then naturally led to use $J_1$ to define an almost complex structure on $T_M|_\Sigma$, and consequently obtain the almost complex structure $\mathbb{J}$ on ${\cal C}$ through the above identifications. Holomorphic  and antiholomorphic tangent vectors in $T_{\cal C}|_{(\Sigma,\calf)}$ are given by (complex) vector fields $ Z$ and $\bar Z$, sections of  $T_M^{\mathbb{C}}|_\Sigma$, satisfying the conditions $Z^m=\frac12(1-iJ_1)_n{}^m Z^n$ and $\bar Z^m=\frac12(1+iJ_1)_n{}^m \bar Z^n$ respectively. From (\ref{null}) and the discussion of section \ref{supsec}  it is clear that the variation of a superpotential with respect to an anti-holomorphic $\bar Z$ vanish identically:
\bea
\bar Z({\cal W})=\frac12\int \bar Z^mP[(\imath_m+g_{mk}dy^k\wedge)\hat\Psi_2]\wedge e^\calf\equiv 0\ .
\eea
Then, the superpotential is holomorphic with respect to the almost complex structure $\mathbb{J}$. Furthermore, it clear that the metric $G$ defined in (\ref{mmetric}) can be identified as a Hermitian metric on ${\cal C}$ naturally inherited from the background metric.

We would like now to argue that, if we restrict to ${\cal C}_{\rm hol}\subset {\cal C}$, the almost complex structure $\mathbb{J}$  reduces to the one naturally induced by the integrable generalized complex structure ${\cal J}_2$ as described above. This can be understood when we  give an interpretation of $\mathbb{J}$ from the point of view of the generalized complex geometry of the internal space $M$.
Suppose to have a subbundle of $(T_M\oplus T_M^*)|_\Sigma$  that can be identified with ${\cal N}_{(\Sigma,\calf)}$ in a particular ``gauge''.  If this subbundle is stable under the action of ${\cal J}_2$ then ${\cal J}_2$ can be used to define an almost complex structure on it. Thus ${\cal J}_2$ induces an almost complex structure  on  ${\cal N}_{(\Sigma,\calf)}$ and as a consequence on ${\cal C}$. The generalized metric structure given by the $SU(3)\times SU(3)$ structure of our backgrounds provides such a subbundle. Let us start by defining the following  orthogonal subspaces of $T_M\oplus T_M^*$ \cite{gualtieri}
\bea
C_\pm={\rm graph}\{\pm g:T_M\rightarrow T_M^* \}\ .
\eea
Note that $C_+\oplus C_-=T_M\oplus T_M^*$ and that $C_\pm$ are both isomorphic to $T_M$ through the projection map $\pi:T_M\oplus T_M^*\rightarrow T_M$. Both $C_+|_\Sigma$ and $C_-|_\Sigma$ indeed provide a subbundle of $(T_M\oplus T_M^*)|_\Sigma$ that is isomorphic to ${\cal N}_{(\Sigma,\calf)}$. In order for them to be  suitable for defining an almost complex structure on ${\cal N}_{(\Sigma,\calf)}$, and then on ${\cal C}$, we have to verify that $C_+|_\Sigma$ and $C_-|_\Sigma$ are stable under the action of ${\cal J}_2$. This can be seen by observing  that, in our $SU(3)\times SU(3)$ structure manifolds, the integrable generalized complex structure ${\cal J}_2$ (and also the non-integrable ${\cal J}_1$) can be written in terms of $J_1$ and $J_2$ by restricting to $C_+$ and $C_-$ and then using the isomorphism $C_\pm\simeq T_M$ \cite{gualtieri}. More precisely, remembering that ${\cal J}_2$ is given by ${\cal J}_+$ in Type IIA and ${\cal J}_-$ in Type IIB, we have that
\bea
{\cal J}_\pm=\pi|_{C_+}^{-1}J_1\pi P_+ \mp \pi|_{C_-}^{-1}J_2\pi P_-\ ,
\eea
where $P_{\pm}$ are the projectors on $C_\pm$.

It is then clear that $C_\pm$ are stable under ${\cal J}_2$ and can be used to define an almost complex structure on ${\cal C}$ as explained above. In particular, using the isomorphism $C_\pm\simeq T_M$, the resulting almost complex structure is essentially given by $J_1$ if we use $C_+$ and by $-J_2$ or $J_2$, in Type IIA or Type IIB respectively, if we use $C_-$. We then see that the choice of $C_+$ is somehow selected by its invariance under mirror symmetry. Indeed, the resulting almost complex structure coincides with the one constructed previously in a more direct way, with respect to which the superpotential is holomorphic. Finally, note that in general the almost complex structure $\mathbb{J}$ defined in this way depends on the $SU(3)\times SU(3)$ structure of the background. However, when  we restrict to ${\cal C}_{\rm hol}$, i.e. to generalized complex submanifolds, this  obviously coincides with the natural one that, as we have already said,  in this case can be defined referring only to the generalized complex structure ${\cal J}_2$.

Consider now the alternative subbundle of $(T_M\oplus T_M^*)|_\Sigma$ isomorphic to $\caln_{(\Sigma,\calf)}$, whose elements are restricted to be of the form $(X_\perp,a)\in\Gamma(T_\Sigma^\perp\oplus T^*_\Sigma)$. Of course any element of $C_+|_\Sigma$ can be put in this form by an appropriate `gauge' transformation. If $(X,g\cdot X)\in C_+$, we can identify it with its image under the translation given by $(-X_{||}, -\imath_{X_{||}}\calf -g\cdot X_\perp)$ (where the meaning of the notation should be obvious). The resulting vector is given by $(X_\perp, (g+\calf)\cdot X_{||})$. Then, using the isomorphism given by $\pi$ to identify $( X,g\cdot X)$ with $ X$, we also find an interpretation from the generalized complex geometry point of view of the identification
\bea                                                                                                                          \label{assos}
 X= X_{||}+ X_\perp\quad\leftrightarrow\quad  (X_\perp, a)\quad {\rm with}\quad a=(g+\calf)\cdot X_{||}\ .
\eea
This identification was already introduced somehow ad hoc in section \ref{supsec} to identify the F-flatness conditions in the form given in (\ref{FDterms}) as conditions for the critical points of the superpotential (\ref{suppot}).

To summarize, we have constructed an almost complex structure $\mathbb{J}$ on the space ${\cal C}$ of the  generalized cycles $(\Sigma,\calf)$,  with respect to which the superpotential ${\cal W}$ defined in (\ref{suppot}) is holomorphic and the metric $G$ defined in (\ref{mmetric}) is Hermitian. At the end of the following section we will see  how this almost complex structure induces also an almost complex structure on the space ${\cal C}_{\rm hol}$ that actually depends only on the  integrable generalized complex structure on $M$.


\section{Reduced configuration and moduli spaces}
\label{complex}

In this section we would like to discuss the gauge symmetries under which the superpotential ${\cal W}$ is left invariant and consider the resulting reduced configuration space and the associated reduced subspace of the space ${\cal C}_{\rm hol}$ of  critical points of ${\cal W}$. Clearly, if we parametrize the possible deformations of the world-volume field-strength $\calf$ with a one-form $a$ as we have explained in section \ref{supsec} (and more extensively in appendix \ref{appA}), then any transformation generated by an exact $a=d\lambda$, with $\lambda$ some function on $\Sigma$, is a gauge symmetry of ${\cal W}$.
Following the previous discussions, the above gauge symmetry generated by $\lambda$ can be identified with the tangent vector field $X_\lambda\in \Gamma(T_\Sigma)$ such that $d\lambda=(P[g]+\calf)\cdot X_\lambda$, that can in turn be seen as a vector tangent to ${\cal C}$ at $(\Sigma,\calf)$. Call ${\bf g}$ the subbundle of $T_{\cal C}$ spanned by such vectors $X_\lambda$.
Now, the holomorphicity of  ${\cal W}$ with respect to the almost complex structure $\mathbb{J}$ automatically implies that  ${\cal W}$ is not only left invariant by the general $X_\lambda$ defined above, but also under its image $\mathbb{J}X_\lambda$ under $\mathbb{J}$, that we can consider as its imaginary extension. This means that ${\cal W}$ is invariant under the action of the general section of the subbundle ${\bf g}^\mathbb{C}$ generated by the vectors of the form $X_\lambda$ and $\mathbb{J}X_\lambda$. Indeed, the holomorphicity of ${\cal W}$ implies that, for any $Y\in \Gamma(T_{\cal C})$,
\bea
(1+i\mathbb{J})Y({\cal W})\equiv 0\ ,
\eea
and then
\bea
X_\lambda({\cal W})\equiv 0 &\Rightarrow& \mathbb{J}X_\lambda({\cal W})\equiv 0\ .
\eea

 It can be clarifying to see how this ``complexification" of the natural $u(1)$ gauge symmetry of the internal generalized cycle
 $(\Sigma,\calf)$ reduces to standard ones when we restrict to the well studied subcases of A and B branes  on  Calabi-Yau 3-folds.  In the Calabi-Yau case, $J_1$ is equal to $J_2$ and is the proper (integrable) complex structure of the Calabi-Yau. Consider first B-branes. These wrap holomorphic cycles with holomorphic connections $A$ on them (such that $\calf=dA$). In this case,  $\mathbb{J}X_\lambda$  generate the transformation $\delta A=i(\partial \lambda-\bar\partial\lambda)$ which is properly identified as an imaginary transformation of the complexified gauge algebra $u(1)^\mathbb{C}=\mathbb{C}^*$. Secondly, consider a Lagrangian A-branes $\Sigma$, with $U(1)$ flat connection A (such that $\calf=dA=0$). In this case $\mathbb{J}X_\lambda$ is associated to a normal vector field  of the form $J_1P[g]^{-1}d\lambda$, which corresponds exactly to the general normal vector field generating Hamiltonian deformations of the Lagrangian A-brane, that must be indeed  considered as gauge symmetries relating equivalent Lagrangians. We then see how our formalism include these specific subcases and provide their natural extension to less trivial ${\cal N}=1$ (and ${\cal N}=2$) flux compactifications.

Note that in the above example with A and B branes, we have really restricted to the space ${\cal C}_{\rm hol}$, while analysis presented above is valid for the whole $\cal{M}$. This has been possible due to the property that $T_{{\cal C}_{\rm hol}}$ is clearly stable under the action of $\mathbb{J}$ and then ${\cal C}_{\rm hol}$ is preserved under the  action of ${\bf g}^\mathbb{C}$ \footnote{The subspace ${\cal C}_{\rm hol}$ can be defined by the condition $d{\cal W}|_{{\cal C}_{\rm hol}}=0$ and then a vector $X\in T_{\cal C}|_{{\cal C}_{\rm hol}}$ tangent to ${\cal C}_{\rm hol}$  can be defined by the condition
\bea                                                                                                                          \label{thol}
d[X({\cal W})]|_{{\cal C}_{\rm hol}}=0\ ,
\eea
where, in each point  $(\Sigma,\calf)\in{\cal C}_{\rm hol}$, we consider $X$ as a  field obtained extending a vector $X\in T_{\cal C}|_{(\Sigma,\calf)}$ to a neighborhood of $(\Sigma,\calf)$ (of course the condition (\ref{thol}) does not depend on the choice of the extension).
From the holomorphicity of the superpotential one can thus conclude that if $X$ is tangent to ${\cal C}_{\rm hol}$ then also $\mathbb{J}X$ is tangent to ${\cal C}_{\rm hol}$. This means that $T_{{\cal C}_{\rm hol}}$ is stable under the action of $\mathbb{J}$. Then, since ${\cal C}_{\rm hol}$ is stable under the action of ${\bf g}$, it is also stable under the action of ${\bf g}^\mathbb{C}$.}.  The subspace ${\cal C}_{\rm hol}$ is also special because, as we have discussed in section \ref{hol},  the almost complex structure $\mathbb{J}$ restricted to it can be defined using only the integrable generalized complex structure on $M$, without any need to involve the $SU(3)\times SU(3)$ structure.
This property  implies  that the generalized complex structure ${\cal J}_2$ on $M$ naturally induces an almost complex structure on ${\cal C}_{\rm hol}$. Note that, since only the integrable generalized complex structure is involved in this definition, all the discussion can be  adapted to the case in which we consider topological branes of the underlying topological model \cite{Kapustin,gmpt2}.

The natural question is if such an almost complex structure on ${\cal C}_{\rm hol}$ is actually integrable. Unfortunately, already in the case of standard Calabi-Yau compactifications the answer is not well understood in general. For example, one can introduce an almost complex structure of the space of Lagrangian submanifolds (with flat $U(1)$ connection) using the symplectic structure of the Calabi-Yau. The resulting almost complex structure mixes embedding and gauge ``coordinates", and its integrability issue is still not clear\footnote{I thank R.~P.~Thomas for correspondence on this point.} (see for example \cite{thomas}). Since our analysis includes this special subcase, we do not try to give an  answer to the problem in the present paper. It would be interesting to understand better this issue from the generalized geometry point of view, that appears to be the natural complex-symplectic unifying language to better approach it.

Finally, observe that one can use $\mathbb{J}$ to naturally induce  an almost complex structure on ${\cal C}^{\rm red}={\cal C}/{\cal G}^\mathbb{C}$, where ${\cal G}^\mathbb{C}$ is the group of finite gauge transformations generated by ${\bf g}^\mathbb{C}$. Furthermore, since the superpotential ${\cal W}$ is left invariant by the action of ${\cal G}^\mathbb{C}$, we can also introduce an almost complex structure on the quotient space ${\cal M}={\cal C}_{\rm hol}/{\cal G}^\mathbb{C}$. As will be clear from the discussion of the following section,  ${\cal M}$ can be identified as the moduli space of the supersymmetric configurations of a space-time filling D-brane. Furthermore, it is known that in the case of  Lagrangian branes on ordinary Calabi-Yau 3-folds, the above almost complex structure on ${\cal C}_{\rm hol}$  descends to an {\em integrable} complex structure on the corresponding ${\cal M}$ (i.e. on the moduli space of special Lagrangian branes). Thus, it seems plausible to hope that the above almost complex structure on ${\cal M}$  can be in fact integrable also in the most general case. We postpone the investigation of this interesting problem to future investigations.

\section{D-flatness and moment map}
\label{symplectic}

In this section we will consider more closely the supersymmetry D-flatness condition written in (\ref{FDterms}). As we already discussed in section \ref{sec2} this condition can be indeed seen as coming from the vanishing of a D-term associated to the effective four-dimensional theory.  As we will now see, the D-flatness condition provide a gauge fixing slice for the action of the imaginary extension of the gauge group, and then select a particular hypersurface ${\cal C}_0$ in ${\cal C}$.
The action of ${\cal G}$ foliates ${\cal C}_0$ in gauge orbits and the base of such a foliation  can be identified as the reduced moduli space ${\cal M}$.

The argument is based on the possibility to see the D-flatness
condition as the vanishing of a moment map associated to the gauge transformation discussed in the previous section, defined with respect to a properly introduced symplectic form. The approach is completely analogous to the one used in the study of branes in Calabi-Yau spaces (see e.g. Chapter 38 of \cite{hori} for a review), even if it differs from it in some details.
Let us start by introducing the following formal symplectic structure on ${\cal C}$. Looking at the vectors $X,Y\in T_{\cal C}|_{(\Sigma,\calf)}$ as sections of $T_{M}|_\Sigma$ by the usual identification, we introduce the following symplectic form
\bea                                                                                                                          \label{sym1}
\Xi (X,Y)|_{(\Sigma,\calf)}=\int_\Sigma X^m Y^n P[e^{2A-\Phi}(\hat\gamma_{mn}+\calf_{mn}){\rm Im}\hat\Psi_1]\wedge
e^\calf\ ,
\eea
where with $\calf_{mn}$ we mean the natural extension with zero orthogonal components of the world-volume field-strength $\calf$ to the complete  $T_M|_\Sigma$, and we recall that the six dimensional gamma matrices $\hat\gamma_m$ act on a form $\omega$ as follows
\bea
\hat\gamma_m\omega=(\imath_m+g_{mn}dy^n\wedge)\omega\ .
\eea
It is easy to see that using the alternative coordinatization  for $T_{\cal C}$ given by $(X_\perp,a)$ and $(Y_\perp,b)$ associated to $X$ and $Y$ respectively by (\ref{assos}),  the above symplectic form takes the form
\bea                                                                                                                          \label{sym1bis}
\Xi [(X_\perp,a),(Y_\perp,b)]|_{(\Sigma,\calf)}&=&\int_\Sigma\Big\{ a\wedge b  \wedge P[e^{2A-\Phi}{\rm Im}\hat\Psi_1]+ P[e^{2A-\Phi}\imath_{X_\perp}\imath_{Y_\perp}{\rm Im}\hat\Psi_1]+\cr
& &+ a\wedge P[e^{2A-\Phi}\imath_{Y_\perp}{\rm Im}\hat\Psi_1] -b\wedge  P[e^{2A-\Phi}\imath_{X_\perp}{\rm Im}\hat\Psi_1]            \Big\}\wedge e^\calf\ .\cr&&
\eea

Note that if we restrict to the case of  D-branes on Calabi-Yau manifolds, the above symplectic structure coincides with the K\"ahler forms constructed for A and B branes.   As in that case, we will not worry whether  $\Xi$  is closed or not, since it will not really be relevant for what follows (for discussions on this point see \cite{thomas}).
Note also  that, in our general case, $\Xi$ cannot be seen as the K\"ahler  form $\Theta$ that can be constructed from the metric $G$  defined in (\ref{mmetric}) and the complex structure $\mathbb{J}$ (i.e. $\Theta(X,Y)=G(X,\mathbb{J}Y)$).
However $\Xi$ and $\Theta$ are related in the following way
\bea                                                                                                                          \label{symrel}
\Theta(X,Y)|_{(\Sigma,\calf)}&=&\frac12\big[\Xi(X,Y)+\Xi(\mathbb{J}X,\mathbb{J}Y) \big]|_{(\Sigma,\calf)}+\cr &&-\frac12\int_\Sigma \big\{\calf(X,Y)+\calf(JX,JY)\big\}P[e^{2A-\Phi}{\rm Im}\hat\Psi_1]\wedge
e^\calf\ .
\eea

We can now introduce the moment map $m: {\cal C}\rightarrow \Gamma(\Lambda^{\rm top}T^*_\Sigma)$ as follows
\bea
m(\Sigma,\calf)=P[e^{2A-\Phi}{\rm Im}\hat\Psi_1]\wedge e^\calf|_{\rm top}\ .
\eea
The moment map $m$ associates    any world-volume function $\lambda$ generating a gauge transformation to the corresponding Hamiltonian function (with respect to the symplectic form $\Xi$) given by the pairing
\bea
\langle m(\Sigma,\calf),\lambda\rangle =\int_{\Sigma} \lambda P[e^{2A-\Phi}{\rm Im}\hat\Psi_1]\wedge e^\calf\ .
\eea
To prove it, it is sufficient to verify that, for any vector $Y\in T_{\cal C}|_{(\Sigma,\calf)}$, we have
\bea
d\langle m(\Sigma,\calf),\lambda\rangle (Y)=\Xi(X_\lambda,Y)\ ,
\eea
where $X_\lambda\in \Gamma(T_\Sigma)$  is the vector generating the gauge transformation and is  defined by the relation $d\lambda=(P[g]+\calf)\cdot X_\lambda$ (see section \ref{complex}).
We can then conclude that the D-flatness condition in (\ref{FDterms}) can be seen as the restriction to the subspace of ${\cal C}$ given by $m^{-1}(0)$. It is clear that any (real) gauge transformation $\lambda$ preserves the constraint $m(\Sigma,\calf)=0$, since for any $h$ we have that
\bea
X_\lambda(\langle m(\Sigma,\calf),h\rangle)=\Xi(X_h,X_\lambda)=\int_\Sigma dh\wedge d\lambda \wedge  P[e^{2A-\Phi}{\rm Im}\hat\Psi_1]\wedge e^\calf \equiv 0\ ,
\eea
where we have used the $d_H$-closedness                                                          of $e^{2A-\Phi}{\rm Im}\hat\Psi_1$.
On the other hand, it is easy to see that  $m^{-1}(0)$ does provide a gauge fixing section for the imaginary gauge transformations. To show this, consider the general imaginary gauge transformation  generated by a vector of the form $\mathbb{J}X_\lambda$, with $\lambda$ a general world-volume function as before. Then one can readily realize that if $(\Sigma,\calf)\in m^{-1}(0)$ then $(\mathbb{J}X_\lambda)(\langle m(\Sigma,\calf),h\rangle)$
cannot vanish for any $h$. To see it, it is enough  to take $h=\lambda$: using the relation (\ref{symrel}) to relate $\Xi$ to the K\"ahler form $\Theta$,
we have that
\bea
(\mathbb{J}X_\lambda)(\langle m(\Sigma,\calf),\lambda\rangle)|_{m^{-1}(0)}&=&\Xi(X_\lambda,\mathbb{J}X_\lambda)|_{m^{-1}(0)}=
\Theta(X_\lambda,\mathbb{J}X_\lambda)|_{m^{-1}(0)}=\cr &=& -G(X_\lambda,X_\lambda)|_{m^{-1}(0)}\ ,
\eea
which generally never vanishes.

Let us note that in the definition of the symplectic structure (\ref{sym1bis}) we have used in an essential way the background metric to again identify ${\cal N}_{(\Sigma,\calf)}$ with $T^\perp_\Sigma\oplus T^*_\Sigma$. This is analogous to what happens in the definition of the almost complex structure $\mathbb{J}$ defined in section \ref{hol}. However, analogously  to what happens that case, it is easy to see that if we restrict to $m^{-1}(0)$ the symplectic structure (\ref{sym1bis}) is canonically defined on sections of  ${\cal N}_{(\Sigma,\calf)}$, in the sense that does not depend on the choice of the subbundle of  $T_M\oplus T^*_M|_\Sigma$ that should represent ${\cal N}_{(\Sigma,\calf)}$ in a particular `gauge'.

Then, to summarize, the D-flatness condition in (\ref{FDterms}) can be  written in the form $m(\Sigma,\calf)=0$ and  clearly provide a global section for the imaginary gauge transformations described in section \ref{complex}. The resulting constrained space $m^{-1}(0)$ is closed under real gauge orbits generated by ${\cal G}$ and the quotient space $m^{-1}(0)/{\cal G}$ provide a characterization of the reduced configuration space ${\cal C}^{\rm red}={\cal C}/{\cal G}^\mathbb{C}$. Furthermore, the same conclusions can be reached if we restrict to the space  ${\cal C}_{\rm hol}$ of generalized complex submanifolds, and then we can make the identifications ${\cal M}={\cal C}_{\rm hol}/{\cal G}^\mathbb{C}=[{\cal C}_{\rm hol}\cap m^{-1}(0)]/{\cal G}$.
Whether in each orbit of ${\cal G}^\mathbb{C}$ inside ${\cal C}_{\rm hol}$ there exists or not a ${\cal G}$ orbit  satisfying the D-flatness condition (and thus minimizing the four-dimensional energy density) can be seen as a generalization of the standard formulation of the stability problem that would be interesting to understand better in the present context.
Finally, we have stressed that (\ref{sym1bis}) can be only formally considered a symplectic form, since it is in general non-closed. However, from the knowledge of what happens in the standard Calabi-Yau case, it is possible to expect that the closedness                                                          can be recovered by restricting to   ${\cal M}$. As the issue of the integrability of the almost complex structure $\mathbb{J}$, the problem to understand in what sense we  can consider  the symplectic structure  (\ref{sym1bis}) as actually closed requires further  investigations.

\section{Examples and applications for D-branes  in \newline
$SU(3)$-structure vacua}\label{examples}

In this section we will consider some basic examples where we can apply explicitly the analysis presented in the previous sections. In particular, we will restrict a little the general setting by focusing on supersymmetric backgrounds with internal $SU(3)$-structure, which are the closest to ordinary flux-less compactifications on Calabi-Yau three-folds. Let us review some of their properties \cite{gmpt,gmpt2,cveticSU3}. The $SU(3)$-structure vacua are characterized by the property that the two internal Weyl spinors $\eta^{(1)}_+$ and $\eta^{(2)}_+$ are actually proportional. It means that we can write them as $\eta^{(1)}=a\eta_+$ and $\eta^{(2)}=b\eta_+$, in terms of a single internal spinor $\eta_+$, such that $\eta_+^\dagger\eta_+=1$. As we have recalled in section \ref{sec1}, since we are considering D-calibrated backgrounds, we must furthermore impose that $|a|=|b|$. Thus we pose
\bea
a=e^{i\varphi_1}|a|\quad,\quad b=e^{i\varphi_2}|a|\ .
\eea
 From $\eta_+$ one can construct an almost complex structure $J$ (with respect to which the internal metric is hermitian) and a $(3,0)$ form $\Omega$ in the following way
\bea
J^m{}_n=-\frac{i}{|a|^2}\eta^\dagger_+\hat{\gamma}^m{}_n\eta\quad,\quad
\Omega_{mnp}=-\frac{i}{a^2}\eta_-^\dagger \hat\gamma_{mnp}\eta_+\ .
\eea
$J$ and $\Omega$ have all the algebraic properties of the complex structure and the holomorphic three form on a standard Calabi-Yau three-fold (see \cite{glmw} for a review). In this case the two normalized pure spinors (\ref{normps}) become
\bea                                                                                                                          \label{pureSU(3)}
\hat\Psi^+=-ie^{i(\varphi_1-\varphi_2)}e^{-iJ}\quad,\quad \hat\Psi^-=-e^{i(\varphi_1+\varphi_2)}\Omega\ .
\eea
Here and in the following we use $J$ to indicate also the K\"ahler form associated to the almost complex structure (the actual meaning being clear from the context). In the particular $\caln=2$ subcase in which the internal space is a standard Calabi-Yau, the expressions (\ref{pureSU(3)}) for the pure spinors are still valid, but with constant arbitrary overall phases. 

From (\ref{pureSU(3)}) it follows that $SU(3)$-structure backgrounds are somehow special in the whole family of $SU(3)\times SU(3)$-structure backgrounds:  in Type IIB the internal space is actually complex with  $c_1(M)=0$ while in Type IIA the internal manifold is symplectic. In the following examples we will re-obtain  these and other needed  properties of the $SU(3)$-structure backgrounds \cite{gmpt,gmpt2}  directly from the supersymmetry conditions (\ref{backsusy}). In this way, we will have a further  case-by-case check of  the deep relation between the ${\cal N}=1$ backgrounds we are considering  and the supersymmetric D-branes they admit. Namely, we will focus on D3, D5, D6 and  D7-branes, with particular attention to this last case. Supersymmetric D4-branes are not allowed in Type IIA $SU(3)$-structure backgrounds \cite{lucal}. D8- and D9-branes can be analyzed along the same lines of the cases explicitly discussed below. Let us make only a comment on the  D9-brane case.
In this case, we can write the  superpotential (\ref{suppot}) by thinking as we had one more dimension. Furthermore the non-abelian generalization is straightforward in this case and simply replaces $\calf\wedge\calf$ with the non-abelian analogous ${\rm Tr}\calf\wedge\calf$. Then, if we consider the case of an internal flux-less Calabi-Yau and $\calf=dA+A\wedge A$, the resulting superpotential becomes up to a constant
\bea
\calw=\frac14\int_{{\rm CY}_3}\Omega\wedge {\rm Tr}(A\wedge \bar\partial A +\frac23 A\wedge A\wedge A)\ ,
\eea
thus reproducing the Witten's Chern-Simons theory describing B-branes filling a Calabi-Yau three-fold \cite{witten92}.

\subsection{D3-branes}
If we consider the simplest case of D3-branes in a $SU(3)$-structure Type IIB background, the space of possible configurations corresponds to the internal space itself, and then has naturally a complex  structure. However,  in general the configuration space is not K\"ahler   without imposing some further condition.

The superpotential (\ref{suppot}) for D3-branes vanishes identically and thus the F-flatness condition is always satisfied. On the other hand, the D-flatness condition is simply given by
\bea                                                                                                                          \label{d3dterm}
\cos(\varphi_1-\varphi_2)|_{y_0}=0\quad\Leftrightarrow \quad(a\pm ib)|_{y_0}=0\ ,
\eea
where $y_0$ is the point of the internal manifolds where the D3-brane is located and the actual sign on the right-hand side of (\ref{d3dterm}) depends on the orientation of the D3-brane. Note that from the first background condition in (\ref{backsusy}) one obtains that $d[e^{2A-\Phi}\cos(\varphi_1-\varphi_2)]=0$ and thus, if the condition (\ref{d3dterm}) is satisfied in a point, it is satisfied everywhere. This is consistent with the fact that in the case of a single D3-brane we do not have any charged matter field under the gauge group and  then we cannot have any non-trivial D-term around a supersymmetric configuration.

The condition $a=\pm i b$ characterizes the so-called type B backgrounds, first considered in \cite{gp,gubser,gp2}, which constitute the supersymmetric subsector of the class of supergravity solutions discussed in \cite{gkp} (see also the recent review \cite{granarev}). In this case, the second condition in (\ref{backsusy}) translates into the following two conditions
\bea                                                                                                                          \label{bsol2}
d\tilde\Omega=0\quad,\quad \tilde\Omega\wedge H=0\ .
\eea
where, to stress the analogy with the standard Calabi-Yau case, we have introduced the holomorphic $(3,0)$-form $\tilde\Omega$ defined as
\bea                                                                                                                          \label{omegaB}
\tilde\Omega=-e^{3A-\Phi}e^{2i\varphi_1}\Omega\ .
\eea
The first condition in (\ref{bsol2}) tells us that all these backgrounds (like all the other $SU(3)$-structure Type IIB vacua) are actually complex. The second condition simply means that $H$ has only $(2,1)$ and $(1,2)$ components.
Looking now at the real part of the first supersymmetry condition in (\ref{backsusy}), we obtain the conditions
\bea                                                                                                                          \label{bsol}
d(e^{2A-\Phi}J)=0\quad,\quad H\wedge J=0\ .
\eea
The first condition implies that the internal space is a warped K\"ahler space, with warp-factor $e^{-2A+\Phi}$, so that the closed K\"ahler form is given by $J^{({\rm K})}=e^{2A-\Phi}J$.  The second condition in (\ref{bsol}), together with the second condition in (\ref{bsol2}), are part of the more general requirement that, in these type B solutions, the complex three form $G_{(3)}=F_{(3)}-\tau H$ (where $\tau=C_{(0)}+ie^{-\Phi}$) must be  $(2,1)$ and primitive.
To obtain the cases in which the internal space is actually a warped Calabi-Yau \cite{gp,gubser}, one must impose that the dilaton is constant (actually $\tau$ must be constant). This condition can be easily obtained by requiring that in the Calabi-Yau case $J^{({\rm K})}\wedge J^{({\rm K})}\wedge J^{({\rm K})}$ must be proportional, up to a constant factor, to $i\tilde\Omega\wedge \bar{\tilde\Omega}$.

From this short review of some of the main properties of the type B vacua, we reach the conclusion that, if the moduli space of supersymmetric D3-branes must coincide with the internal manifold itself, then it is automatically a K\"alher manifold that, if we furthermore require a constant dilaton, is also Calabi-Yau.

\subsection{D5-branes}
In the case of D5-branes wrapping an internal two-cycle $\Sigma$ the superpotential (\ref{suppot}) takes the form of the Witten's superpotential \cite{wittenQCD}
\bea
{\cal W}={\cal W}_0+\frac12\int_{{\cal B}}P[\tilde\Omega]\ ,
\eea
where again we used the holomorphic $(3,0)$-form  $\tilde\Omega\equiv e^{3A-\Phi}\hat\Psi^-$. By considering a complex coordinatization $z^i$ ($i=1,2,3$) of the internal space, we can specify the embedding using
complex fields $\phi^i(\sigma)$ (where $\sigma^\alpha$ are world-volume coordinates). The superpotential is clearly  holomorphic with respect to these complex fields. However, if we want to consider the superpotential as a functional on the space of diffeomorphism equivalent cycles, the background complex structure does not naturally induce a complex structure for it, and we need some additional structure.  Indeed, the almost complex structure  $\mathbb{J}$ introduced in section \ref{hol} uses two additional ingredients: the background metric that allows to identify explicitly the normal bundle of the two-cycle with its orthogonal bundle, and the world-volume gauge field, which is in general mixed with the embedding coordinates under the action of the almost complex structure.


Turning to the D-flatness condition, it takes the form
\bea                                                                                                                          \label{d5dflat}
\calf=-{\rm tg}(\varphi_1-\varphi_2)P_\Sigma[J]\ .
\eea
Then, if one wants to admit supersymmetric D5-branes with zero $\calf$, it is natural to impose everywhere the condition $\varphi_1-\varphi_2=0$ or $\pi$, or equivalently $a=\pm b$ (more in general, it would be sufficient to impose such a condition only where the brane is located). This condition defines the so called type C backgrounds (see e.g. \cite{granarev} for more on them), of which the solution found in \cite{volkov}, and interpreted as a background dual to a confining gauge theory in \cite{mn}, provides the most known explicit example. A different and somehow special case is obtained by considering a type B background, that is $\varphi_1-\varphi_2=\pm \frac\pi2$. The D5-brane can then be supersymmetric only if it wraps a collapsed cycle (so that $P_\Sigma[J]=0$) with a non-vanishing $\calf$ field on it, in such a way to have a non-vanishing tension. The resulting configurations are fractional D3-branes, that are well known supersymmetric configurations giving rise to corresponding backgrounds with fluxes.

Note that, in the case of type C solutions  (fixing for example $\varphi_1-\varphi_2=0$), from the real part of the first  supersymmetry  condition in (\ref{backsusy}) one  can directly obtain the conditions $d(2A-\Phi)=0$ and $H=0$. Thus, it is not difficult to see that the symplectic form (\ref{sym1bis}) takes the form
\bea
\Xi(X,Y)=-e^{2A-\Phi}\int_\Sigma\big\{ a\wedge b+ \frac12 P[\imath_{Y_\perp}\imath_{X_\perp}(J\wedge J)]\big\}
\eea
where we have moved $e^{2A-\Phi}$ out of the integral since it is constant.
If moreover we restrict to the case of holomorphic two-cycles, this symplectic form becomes
\bea                                                                                                                          \label{restrsymd5}
\Xi(X,Y)=-e^{2A-\Phi}\int_\Sigma\big\{ a\wedge b+J(X_\perp,Y_\perp)P[J]\big\}\ .
\eea
One can immediately check that the moment map for the ordinary gauge transformations $a=d\lambda$ is given by $m(\Sigma,\calf)=-e^{2A-\Phi}\calf$, consistent with our general discussion. Imposing that it must vanish is of course equivalent to the D-flatness condition (\ref{d5dflat}).
Note that the restricted $\Xi$ given in (\ref{restrsymd5}) can be directly related to the K\"ahler two-from $\Theta$ (see section \ref{symplectic}) in the following way
\bea
\Xi(X,Y)=\Theta(X,Y)-e^{2A-\Phi}\int_\Sigma \calf(X,Y)\calf \ .
\eea
Note that, also comparing with the general formula (\ref{symrel}), in this case we have clearly that $\Xi(X,Y)$ is of the type $(1,1)$. Thus, it can be seen as a deformation of the K\"ahler form $\Theta$ due to the presence of nontrivial  world-volume $\calf$.

\subsection{D6-branes}
Let us now consider the case of a D6-brane wrapping an internal three-cycle in a Type IIA $SU(3)$-structure background. Note first of all that, from the second supersymmetry condition in (\ref{backsusy}), we immediately obtain that $3A-\Phi +i(\varphi_1-\varphi_2)$ must be constant, $H$ must vanish, and $dJ=0$. This explicitly checks the known property that the internal space must be symplectic (but in general not complex). Thus, we can write the superpotential for D6-branes in the following explicit form
\bea                                                                                                                          \label{suppotlagr}
{\cal W}={\cal W}_0-\frac12 e^{3A-\Phi+i(\varphi_1-\varphi_2)}\int_{\cal B}\big\{ P[J]\wedge \tilde\calf +\frac i2 P[J\wedge J]-\frac i2 \tilde\calf\wedge\tilde\calf\big\}\
\eea
 One can easily check that in this case a generalized complex three-cycle corresponds to a Lagrangian submanifold with vanishing field-strength, i.e.
\bea
P_\Sigma[J]=0\quad,\quad \calf=0\ .
\eea
Note that (\ref{suppotlagr}) is completely identical in form to the holomorphic functional presented for example in \cite{thomas}, that can be written in the form of the standard Chern-Simons action that was proved in \cite{witten92} to describe Lagrangian A-branes.

Let us now consider the D-flatness condition, which reads
\bea
P_\Sigma[{\rm Im}\tilde\Omega]=0\ .
\eea
where again we have posed $\tilde\Omega =e^{2A-\Phi}e^{i(\varphi_1+\varphi_2)}\Omega$, which obeys the condition $d({\rm Im}\tilde\Omega)=0$.
Note that in this case we do not have any immediate  constraint to be imposed on the background in order for it to admit a supersymmetric D6-brane. The only obvious necessary condition is the following topological condition that must be imposed on the brane
\bea
\int P_\Sigma[{\rm Im}\tilde\Omega]=0\ .
\eea

The  formal symplectic form (\ref{sym1bis}), when evaluated in a general point of the configuration space ${\cal C}$, is explicitly given by
\bea
\Xi(X,Y)=\int_\Sigma\big\{a\wedge P[\imath_{Y_\perp}{\rm Im}\tilde\Omega]-b\wedge P[\imath_{X_\perp}{\rm Im}\tilde\Omega] +P[\imath_{X_\perp}\imath_{Y_\perp}{\rm Im}\tilde\Omega]\wedge\calf \big\}\ .
\eea
If we now restrict to the superpotential critical subspace ${\cal C}_{\rm hol}$ of Lagrangian cycles, the symplectic form reduces to
\bea
\Xi(X,Y)=\int_\Sigma\big\{a\wedge P[\imath_{Y_\perp}{\rm Im}\tilde\Omega]-b\wedge P[\imath_{X_\perp}{\rm Im}\tilde\Omega]\big\}\ .
\eea
This is identical in form to the  symplectic form introduced in \cite{thomas} for Lagrangian branes with flat $U(1)$ connection on ordinary Calabi-Yau three-folds.

\subsection{D7-branes}

The final case that we analyze explicitly is that of a D7-brane in a Type II $SU(3)$-structure background. In order to be more concrete, let us focus on the case in which the background is of the type B described in the discussion about D3-branes. As we have already said, these backgrounds  can be thought of as generated by D3 and/or fractional D3 and/or D7-branes \cite{gp2} and their internal manifold has a warped K\"ahler metric with  K\"ahler form $J^{({\rm K})}=e^{2A-\Phi}J$ and  a global holomorphic $(3,0)$-form $\tilde\Omega$ defined in (\ref{omegaB}). This will allow us to discuss some interesting additional issues related to the moduli space of the D7-brane.

In type B backgrounds, the superpotential for the D7-brane is given by
\bea                                                                                                                          \label{d7}
{\cal W}(\Sigma,\calf)={\cal W}_0+\frac12\int_{\cal B} P[\tilde\Omega]\wedge \tilde\calf\ .
\eea
We already know that the extrema of this superpotential are given by D-branes wrapping generalized complex submanifolds $(\Sigma,\calf)$, with $\Sigma$ holomorphically embedded and $\calf$ of kind $(1,1)$. Let us see this directly from the superpotential (\ref{d7}). The variation with respect of the world-volume gauge field gives the condition
\bea                                                                                                                          \label{eqmoto1}
P_\Sigma[\tilde\Omega]=0\ .
\eea
This condition requires the cycle $\Sigma$ to be holomorphically embedded. The additional F-flatness condition that $\calf$ must be of kind $(1,1)$ can be derived by varying the embedding coordinates along the vector field $X\in\Gamma(T_M|_\Sigma)$ and transforming the world-volume field strength accordingly to the rule $\delta \calf=P_\Sigma[\imath_X H]$. The resulting derivative of the superpotential is given by
\bea                                                                                                                          \label{eqmoto2}
P_\Sigma[\imath_X\tilde\Omega]\wedge \calf\ ,
\eea
which clearly vanishes only if $\calf_{(0,2)}=0$. Note that, as we have already discussed in general in section \ref{supsec}, once we take into account the other condition (\ref{eqmoto1}), the condition (\ref{eqmoto2}) is well defined also thinking to $X$ as a  section of the canonical normal bundle ${\cal N}_\Sigma=T_M|_\Sigma/T_\Sigma$.

Now, in principle the superpotential (\ref{d7}) takes into account all the possible internal fluctuation modes of the D7-brane. One could then `integrate out' the heavy massive modes, to obtain an effective superpotential for the light ones, as described in \cite{witten92}. More directly,  the superpotential (\ref{d7}) allows to immediatly discuss a mechanism of flux-generated lifting of the possible moduli fields corresponding to the infinitesimal deformations of a holomorphic cycle. Such an effect was discussed in \cite{marchesano} using a different procedure in the less general case where the internal manifold is a warped Calabi-Yau. The following discussion generalizes it and clarifies its origin.  

Let us first recall that the possible infinitesimal deformations of a holomorphic cycle $\Sigma$ (of arbitrary dimension) are given by the space of global sections $H^0(\Sigma,{\cal N}^{\rm hol}_\Sigma)$ of the holomorphic normal bundle ${\cal N}^{\rm hol}_\Sigma=T_M^{1,0}|_\Sigma/T_\Sigma^{1,0}$. In our case, $\Sigma$ is  a divisor. By the triviality of the canonical bundle of $M$ and the adjunction formula, one obtains the standard result that $H^0(\Sigma,{\cal N}^{\rm hol}_\Sigma)=H^{2,0}(\Sigma)$. Thus, there are  $h^{2,0}(\Sigma)={\rm dim}H^{2,0}(\Sigma)$ possible moduli deformations parametrized by complex coordinates $t^i$, $i=1,\ldots,h^{2,0}(\Sigma)$.

The first order derivative by $t^i$ of  the superpotential (\ref{d7}) is given by 
\bea                                                                                                                          \label{first}
{\partial_i \cal W}=\frac12\int_\Sigma P_{\Sigma}[\imath_{X_i}\tilde\Omega]\wedge \calf\ ,
\eea
where $X_i$ is the holomorphic section of ${\cal N}^{\rm hol}_\Sigma$ generating the shift in $t^i$. Obviously (\ref{first}) vanishes in a  point $t_0$ where $\calf$ is $(1,1)$ and in general one obtains a set of $h^{2,0}(\Sigma)$ possible moduli lifting conditions
\bea\label{second}
a_i(t)\equiv\int_\Sigma P_{\Sigma}[\imath_{X_i}\tilde\Omega]\wedge \calf=0\ ,
\eea
that can in principle lift all the possible $h^{2,0}(\Sigma)$ moduli fields $t^i$.
In \cite{marchesano},   the set of $h^{2,0}(\Sigma)$ conditions $a_i(t)=0$ were found by a rather different way in the warped Calabi-Yau subcase,  conjecturing that the $a_i$'s could be identified as the first derivatives of a superpotential. Equation (\ref{first}) gives a direct confirmation and generalization of that proposal.

Possible holomorphic mass terms can be now in principle computed by taking a further derivative of the superpotential. Let us first of all recall  that, already in the flux-less Calabi-Yau case,  when $T_M^{1,0}|_\Sigma$ does not holomorphically split into $T_\Sigma^{1,0}\oplus \caln^{\rm hol}_\Sigma$ some of the  $h^{2,0}(\Sigma)$ infinitesimal  embedding deformations may be in fact massive, due to possible obstructions coming from the holomorphic line bundle on the brane \cite{katz}.  The superpotential (\ref{d7}) directly exhibits the possible presence of this kind of obstructions, even in the more general case of backgrounds with fluxes we are considering. Indeed,  the variation of  $P_{\Sigma}[\imath_{X_i}\tilde\Omega]$ in (\ref{first})  may produce in general a $(1,1)$ form that, combined with a non-trivial $(1,1)$ world-volume field-strength  $\calf$, can give  non-vanishing mass terms for the $t^i$'s \footnote{This observation is due to F.~Denef, who I thank for discussions on this point.}.

In order to focus on  mass terms that are a peculiar effect of the background fluxes, let us now assume the holomorphic splitting of   $T_M^{1,0}|_\Sigma$ into $T_\Sigma^{1,0}\oplus \caln^{\rm hol}_\Sigma$. In this case, when the internal manifold is a standard flux-less Calabi-Yau, the  $t^i$'s are massless, even if there can be  possible higher order obstructions   (like the standard ones that lies in  $H^1(\Sigma,{\cal N}^{\rm hol}_\Sigma)$  \cite{koda}) that  should be described by  non-trivial higher order  terms in the superpotential (\ref{d7}) (see for example the related discussions in \cite{quintic,kachru,agana}). However, from the   superpotential (\ref{suppot}) one can easily realize that, in presence of a non-trivial background $H$-flux, the $t^i$'s can in general cease to be massless. 
To see this, it is enough to take the second derivative of the superpotential around a point $t_0$ corresponding to  a  generalized holomorphic cycle $(\Sigma_0,\calf_0)$,  obtaining the following holomorphic mass matrix
\bea                                                                                                                          \label{holmass}
m_{ij}(t_0)\equiv (\partial_i\partial_j{\cal W})(t_0)=\frac12\int_{\Sigma_0} P_{\Sigma_0}[\imath_{X_i}\tilde\Omega
\wedge\imath_{X_j}H]\ .
\eea 
The formula (\ref{holmass}) explicitly shows how a nontrivial $H$-flux can induce holomorphic mass terms (that would otherwise vanish) for the possible embedding moduli.

If the F-flatness conditions are satisfied, in order to obtain a full supersymmetric configuration we have still to impose the D-flatness condition, that in this case reads
\bea                                                                                                                          \label{Dd7}
P_\Sigma[J]\wedge\calf=0\ .
\eea
Such a condition is a generalization of what is known as a Hermitian-Yang-Mills condition in standard Yang-Mills theories. We can then easily adapt the standard argument for an abelian Yang-Mills theory (see for example \cite{GSW}) to prove that in each orbit of different $\calf$'s generated  by the imaginary extension of the (abelian) gauge group there is a particular $\calf$ satisfying the D-term condition (\ref{Dd7}) if and only if the condition
\bea                                                                                                                          \label{topd7}
\int_\Sigma P[e^{2A-\Phi}J]\wedge \calf=0\ ,
\eea
is satisfied. Indeed the imaginary gauge transformation acts as $\delta\calf=i\partial\bar\partial\lambda$, where $\lambda$ is any real function on $\Sigma$ and $\partial$ and $\bar\partial$ are the standard Dolbeault differential operators on $\Sigma$. It immediately follows that the condition (\ref{topd7}) is necessary and sufficient for the existence of a $\lambda$ such that the transformed $\calf$ satisfies (\ref{Dd7}). Note also that the condition (\ref{topd7}) is actually topological (in the sense that it is left invariant by any continuous deformation of $\Sigma$ and $\calf$), due to  the primitivity condition $J\wedge H=0$. 

Also, from the general discussion of section \ref{symplectic}, we know that the D-flatness condition (\ref{topd7}) can be obtained as the vanishing moment map condition associated to the symplectic form (\ref{sym1bis}). 
Restricting to supersymmetric configurations, for which $\calf=-*_4\calf$, we can rewrite it  in the form 
\bea                                                                                                                          \label{sympld7bis2}
\Xi[(X_i,a),(\bar X_{\bar k},\bar b)]|_{\rm susy}&=&-\int_\Sigma e^{2A-\Phi} a\wedge \bar b\wedge P[J]+\cr
&& -\frac i8\int_\Sigma e^{2A-\Phi}(1+\frac12 \calf^2)P[\imath_{X_i}\Omega\wedge \imath_{\bar X_{\bar k}}\bar \Omega] \ ,
\eea
where the indexes in $\calf^2$ are contracted with the induced metric $P[g]$. 

As a further application, from the  formulas (\ref{mmetric}) and  (\ref{pertpot2})  we can also find a general formula for the flux-induced physical mass term for the embedding moduli $t^i$ around a supersymmetric configuration $(\Sigma_0,\calf_0)$. From (\ref{mmetric}),  one immediately obtains that the  metric for the embedding holomorphic deformations  is given by
\bea                                                                                                                          \label{intrmetric}
G_{i\bar k}=\frac 18\int_{\Sigma} e^{2A-\Phi}(1+\frac12 \calf^2)P_{\Sigma_0}[\imath_{X_i}\Omega\wedge \imath_{\bar X_{\bar k}}\bar \Omega] \ .
\eea
In the approximation in which the warp-factor, the non-trivial dilaton and $\calf$ can be neglected, this metric reduces to the K\"ahler metric for the embedding moduli found in \cite{louis1}.
Thus, from (\ref{pertpot2}) and reintroducing the tension as described after (\ref{cansuppot}), we obtain the following quadratic term in the potential
\bea
{\cal V}\simeq M^2_{i\bar k}(t-t_0)^i (\bar t-\bar t_0)^{\bar k} +\ldots\ ,
\eea
where the physical mass matrix $M^2$ is given by
\bea
M_{i\bar k}^2= G^{r  \bar s}(t_0)m_{i r}(t_0)\bar m_{\bar k \bar s}(t_0)\ .
\eea
Thus, the superpotential generates flux-induced mass terms for the embedding moduli of the D7-branes in a general type B background.  Analogous massive terms where computed  in \cite{marchesano} by a  different  argument in the subcase of internal warped Calabi-Yau spaces and $\calf_0=0$.

To be even more concrete, we could consider the simplified case when the divisor $\Sigma$ has trivial canonical bundle and we can write $M\simeq\Sigma\times \mathbb{C}$ globally on a cylindrical neighborhood of $\Sigma_0$, in such a way that $J^{({\rm K})}=P_\Sigma[J^{({\rm K})}]+idt\wedge d\bar t$, where $t$ is the holomorphic transverse coordinate in $\mathbb{C}$. In this case the only holomorphic embedding deformation given by the position $t$ of $\Sigma$ in $\mathbb{C}$. Let us introduce the holomorphic $(2,0)$ form $\omega$ on $\Sigma$ such that $\tilde\Omega=\omega\wedge dt$. Then we have that
\bea
G_{t\bar t}&= &\int_\Sigma e^{\Phi}(1+\frac12 \calf^2)dVol_4\ ,\cr
m_{tt}&=&-\frac{ig_s}{4}\int_\Sigma e^{\Phi}\omega\wedge P[\imath_{\partial_t}\bar G_{(3)}]\ .
\eea
Thus, by posing $t=2\pi\alpha^\prime\phi$ and $g^2_{YM}=(2\pi)^5(\alpha^\prime)^2g_s/G_{t\bar t}$, we can write the following canonical quadratic Lagrangian for $\phi$
\bea                                                                                                                          \label{quadrd7}
{\cal L}=-\frac{1}{g^2_{YM}}(\partial_\mu\phi\partial^\mu\bar\phi +M^2\phi\bar\phi)\ .
\eea
where $M^2=G_{t\bar t}^{-2}|m_{tt}|^2$. As a check, we can consider the simplest case where the internal manifold is a six torus $(\mathbb{T}^2)^3$, and assume that  the warped factor and the $G_{(3)}$ are constant on the wrapped $\Sigma=(\mathbb{T}^2)^2$. In this case, defining $S=2G_{12\bar t}$, we can write
\bea                                                                                                                          \label{susymass}
g^2_{YM}=\frac{(2\pi)^5(\alpha^\prime)^2g_s}{Vol_4(\Sigma)-\frac12\int_\Sigma\calf\wedge\calf}\quad,\quad  M^2= \frac{g_s^2e^{8 A_0}|S|^2}{8+2\left(\frac{\int_\Sigma\calf\wedge\calf}{Vol_4(\Sigma)}\right)^2}\ .\eea
The result (\ref{susymass}) provides the generalization to arbitrary  $\calf\neq 0$ of the supersymmetric massive term found in \cite{softuranga} in the case $\calf=0$ by direct dimensional reduction of the D7-brane action.

Finally, if we further assume that we can make a gauge choice such that the $B$ field is a globally defined $(1,1)$-form  on  $M\simeq\Sigma\times \mathbb{C}$, then the superpotential (\ref{d7}) can be written in the form
\bea\label{d7bis}
{\cal W}=\frac12 \int_\Sigma t\omega\wedge f\ ,
\eea
where $f=\calf -P[B]$ is the proper $U(1)$ field strength on $\Sigma$. This superpotential coincides with the superpotential found in \cite{lust2}, for a class of F-theory backgrounds, and in \cite{louis2}, by dimensional reduction of the DBI plus CS D7-branes action on  Calabi-Yau orientifolds. In these papers was also found a D-term of the form
\bea\label{lustd7}
D\sim \int_\Sigma P[J^{({\rm K})}]\wedge\calf \ ,
\eea
where $\calf$ was assumed to be harmonic. This same form can be found from our D-term
\bea\label{ourdd7}
{\cal D}d\sigma^1\wedge\ldots\wedge=P_{\Sigma}[e^{2A-\Phi}J]\wedge\calf\ ,
\eea
by simply noting that, expanding $\calf$ in a base of harmonic forms, in (\ref{ourdd7}) only the non-primitive component (proportional to $P_\Sigma[e^{2A-\Phi}J]$) of $\calf$ survives and its contribution  to our D-term is essentially given by (\ref{lustd7}).

\section{Concluding remarks}\label{discussion}

In this paper we have approached the problem of giving  a unified description of the dynamics of a general  D-brane on a general ${\cal N}=1$ background. In particular we have identified the F- and D-terms of the corresponding supersymmetric four-dimensional description. By introducing an appropriate metric on the configuration space, we have also shown how the resulting four-dimensional potential around a supersymmetric configuration can be written in the standard form dictated  by  $\caln=1$ supersymmetry. Furthermore we have seen how the corresponding F-flatness conditions can be derived from a superpotential that can be expressed in a universal way by using the integrable pure-spinor \cite{gmpt2} of the underlying space, while the D-flatness condition can be seen as the vanishing of a moment map whose definition involves the {\em non}-integrable pure spinors.

It was possible to take  the analysis on very general grounds  thanks to the generalized calibrations introduced in \cite{lucal}. They have not only simplified many technical steps but they have also provided an elegant  geometrical interpretation of the resulting supersymmetric structure, due to their relation to the possible solitonic objects of the four-dimensional theory obtained through their D-brane realization. For example, in section \ref{supDW} we have seen how the form of the superpotential ${\cal W}$ presented in (\ref{suppot}) can be immediately guessed by using   the generalized calibration $\omega^{\rm (DW)}$ in (\ref{others}) for domain wall D-branes   and the well known relation between the superpotentials and BPS domain walls. The argument is  completely analogous to the one used in \cite{gukov1,gukov2} to find effective closed string superpotentials, and also provides  a non-trivial  consistency check of our results.

Regarding the D-terms, we have discussed in section \ref{Dstrings} how they can be related to  cosmic strings, which constitute the other possible BPS solitonic objects allowed by the effective $\caln=1$ four-dimensional theory. In particular, using the generalized calibration $\omega^{\rm (string)}$ written in (\ref{others}), we have exactly reproduced from a purely D-brane setting the cosmic string tension obtained from effective four-dimensional arguments in $\cite{toinestring}$. This gives a strong explicit check not only of the correspondence proposed in \cite{toinestring} between supergravity cosmic strings and  cosmic strings obtained by wrapping D-branes on internal cycles, but also of the interpretation of the supergravity theory the authors of  \cite{toinestring} started from as a good effective four-dimensional theory describing a brane-antibrane system coupled to the closed string sector (see also the discussion in \cite{dvali}).

If on one side the supersymmetric solitons constructed from D-branes  provide a physical interpretation of the effective $\caln=1$ structure presented in this theory, on  the other side a proper understanding of the underlying mathematical structure seems to require some more effort. We have proposed some first results in this direction, by presenting an almost complex structure and a symplectic structure on the configuration space that are naturally associated to the superpotentials and the D-terms of the four-dimensional description. However, we have worked at the formal level and a deeper mathematical control of these structures would be desirable. First of all, the above structures are not trivially integrable. This is somehow expected, since the same happens even in the simpler case of branes on Calabi-Yau spaces \cite{thomas}. However, in that case, restricting to the moduli space of supersymmetric branes the integrability of the complex and symplectic structure is recovered, moreover obtaining a  resulting K\"ahler structure. This is compatible with the $\caln=1$ supersymmetry and in our more general case we then expect something similar when we really restrict to the moduli space of the supersymmetric configurations. This would require a better understanding of the moduli space of the generalized calibrated submanifolds of \cite{lucal}. In any case, as \cite{lucal} and this paper show, generalized complex geometry  seems to be the right language to properly  address these problems in a unified way.

The generality of the whole discussion automatically implies also the complete symmetry of the results if we pass from  Type IIA to Type IIB (and vice-versa) and contemporary exchange the two pure spinors $\Psi^\pm$. This can be seen as a formal generalized mirror symmetry relating $\caln=1$ flux backgrounds \cite{gmpt2}, and it would be very interesting to try to give some more substantial arguments in favor of it (for discussions on generalized mirror symmetry see e.g. \cite{glmw,Fidanza,Ben-Bassat,Jeschek,Grange,Tomasiello,Grana}). For example, it would be interesting to address the problem starting from a SYZ approach \cite{syz}, where D-branes play a central role and then our analysis could be helpful.

Finally, even if we have mainly focused on the purely theoretical aspects, we hope that the results could be useful also in concrete constructions of always more realistic models in string theory that have flux compactifications and D-branes as the essential ingredients (like for example in the KKLT proposal \cite{kklt}). The explicit study of D7-branes on $SU(3)$-structure backgrounds presented in section \ref{examples} provides an example of how our  analysis allows to reproduce and generalize some previous results in that direction.

\vskip 1.3cm

\begin{center}
{\large  {\em Acknowledgments}}
\end{center}
I would like to thank R.~Argurio, A.~Celi, F.~Denef, F.~Ferrari, M.~Gra\~na, R.~Minasian,  M.~Petrini, P.~Smyth, R.~P.~Thomas , J.~Van den Bergh and  A.~Van Proeyen   for useful discussions.  This work is supported in part by the Federal Office for Scientific, Technical and Cultural Affairs through the "Interuniversity
Attraction Poles Programme -- Belgian Science Policy" P5/27 and by
the European Community's Human Potential Programme under contract
MRTN-CT-2004-005104 `Constituents, fundamental forces and
symmetries of the universe'.

\vskip 2cm


\begin{appendix}

\section{Deformations of  D-branes on non-trivial $B$ field}\label{appA}

In order to better understand the possible infinitesimal deformations of the the generalized cycle $(\Sigma,\calf)$, let us briefly review the definition of twisted world-volume gauge field in the presence on a non-trivial $B$ field \cite{freedwitten,kapuB} on the internal manifold $M$.

A non-trivial $B$ field can be seen as a connection of a gerbe on $M$ \cite{HitchinL}. Consider an open covering $\{ \calu_\alpha\}$ of M. Then a {\em gerbe} is defined by a \u{C}ech cocycle $\{g_{\alpha\beta\gamma}\}$ of maps $g_{\alpha\beta\gamma}:\calu_\alpha\cap \calu_\beta\cap \calu_\gamma\rightarrow U(1)$ (the cocycle condition is given by the condition $g_{\beta\gamma\delta}g_{\alpha\gamma\delta}^{-1}g_{\alpha\beta\delta}g_{\alpha\beta\gamma}^{-1}=1$ on any $\calu_\alpha\cap \calu_\beta\cap \calu_\gamma\cap \calu_\delta\neq \emptyset $). The inequivalent gerbes are then defined by elements of the second \u Cech cohomology group $\check H(M,C^\infty(U(1)))\simeq H^3(M,\mathbb{Z})$. In string theory the $B$ field defines a {\em connection} on a gerbe. Namely, we can take an open covering $\{\calu_\alpha\}$ such that the $B$ field is locally given by a two form $B_\alpha$ on any $\calu_\alpha$. Then, on any twofold intersection $\calu_\alpha\cap\calu_\beta$ there are one-forms $\Lambda_{\alpha\beta}$ such that
\bea                                                                                                                          \label{gerbeconn}
B_\alpha-B_\beta&=& 2\pi\alpha^\prime d\Lambda_{\alpha\beta}\quad\quad\quad\quad\,\,\quad  {\rm on}\quad \calu_\alpha\cap\calu_\beta\ ,\cr
\Lambda_{\alpha\beta}+\Lambda_{\beta\gamma}+\Lambda_{\gamma\alpha}&=&-ig_{\alpha\beta\gamma}^{-1}dg_{\alpha\beta\gamma}\quad\quad{\rm on}\quad\calu_\alpha\cap\calu_\beta\cap\calu_\gamma\ .
\eea
The globally defined three-form $H=dB$ is normalized in such a way  that $[H/(2\pi)^2\alpha^\prime]\in H^3(M,\mathbb{R})$ is the image of an integral class in $H^3(M,\mathbb{Z})$ and represents in real cohomology the characteristic class of the gerbe.

In presence of such a gerbe with connection,  for a D-branes wrapping a submanifold $\Sigma$, we can take an open covering $\tilde\calu_\alpha=\Sigma\cap\calu_\alpha$ on $\Sigma$. Then, a  ``$U(1)$ connection"  on the D-brane is given by a set of one-forms $A_\alpha$ defined on $\tilde\calu_\alpha$ and a set of transition functions $h_{\alpha\beta}:\tilde\calu_\alpha\cap\tilde\calu_\beta\rightarrow U(1)$ such that
\bea                                                                                                                          \label{wwgauge}
A_\alpha-A_\beta+ih_{\alpha\beta}^{-1}dh_{\alpha\beta}&=&P_\Sigma[\Lambda_{\alpha\beta}]\quad\quad\ \ {\rm on}\quad\tilde\calu_\alpha\cap\tilde\calu_\beta\ ,\cr
h_{\alpha\beta}h_{\beta\gamma}h_{\gamma\alpha}&=&P_\Sigma[g_{\alpha\beta\gamma}]\quad\quad\ {\rm on}\quad\tilde\calu_\alpha\cap\tilde\calu_\beta\cap\tilde\calu_\gamma\ .
\eea
The world-volume globally defined field strength is given by $\calf=2\pi\alpha^\prime dA_\alpha+P_\Sigma[B_\alpha]$, and obeys the modified Bianchi identity $d\calf=P_\Sigma[H]$.

Now, consider any other $U(1)$ connection $A_\alpha^\prime$ with transition functions $h_{\alpha\beta}^\prime$, on the same cycle $\Sigma$. Then the set of one-forms $a_\alpha/2\pi\alpha^\prime=A^\prime_\alpha-A_\alpha$  define a proper connection on the line bundle on $\Sigma$ defined by the transitions functions $g_{\alpha\beta}=h_{\alpha\beta}^\prime h_{\alpha\beta}^{-1}$ (such that $g_{\alpha\beta}g_{\beta\gamma}g_{\gamma\alpha}=1$). If we consider an infinitesimal deformation of the (twisted) $U(1)$ connection $A$, it is described by a globally defined 1-form $a$ on $\Sigma$, which can be seen as a connection on the trivial line bundle on $\Sigma$. The corresponding infinitesimal deformation of the world-volume field strength is given by $\delta\calf=da$.

Till now we have kept fixed the cycle $\Sigma$ wrapped by the brane. However, we can consider also a deformation of it, generated by a vector field $X\in\Gamma(T_N)$, where $N\subset M$ is an open neighborhood of $\Sigma$.  Obviously, under such a deformation, the background gerbe transition functions $g_{\alpha\beta\gamma}$ are deformed to new $g^\prime_{\alpha\beta\gamma}\simeq g_{\alpha\beta\gamma}+\call_X g_{\alpha\beta\gamma}$, together with a new gerbe connection defined by $B^\prime_\alpha\simeq B_\alpha+\call_X B_\alpha$ and $\Lambda^\prime_{\alpha\beta}\simeq \Lambda_{\alpha\beta} +\call_X \Lambda_{\alpha\beta}$.  It is clear that also the transition functions $h_{\alpha\beta}$ which enter the definition of world-volume gauge connection in (\ref{wwgauge}) must transform accordingly. Using (\ref{gerbeconn}) it is not difficult to see that $h^\prime_{\alpha\beta}\simeq h_{\alpha\beta}(1+iP_\Sigma[\imath_X\Lambda_{\alpha\beta}])$. Thus, also the gauge field $A$ must be deformed to $A_\alpha^\prime=A_\alpha-P_\Sigma[\imath_X B_\alpha]/2\pi\alpha^\prime$. Note that the new gerbe defined by the transition functions $\{g_{\alpha\beta\gamma}^\prime\}$  is related to the gerbe defined by $\{ g_{\alpha\beta\gamma}\}$ through the ``gauge transformation" $g^\prime_{\alpha\beta\gamma}=g_{\alpha\beta\gamma}(f_{\alpha\beta}f_{\beta\gamma}f_{\gamma\alpha})$, where $f_{\alpha\beta}\simeq 1+i\imath_X\Lambda_{\alpha\beta}$. We can then perform a further gauge transformation $B_\alpha^\prime\rightarrow \tilde B_\alpha=B^\prime_\alpha- d(\imath_X B_\alpha)\simeq B_\alpha+\imath_X H$, obtaining a new connection defined by the new $\tilde B_\alpha$'s for the undeformed gerbe defined by the transition functions $g_{\alpha\beta\gamma}$ and with the same transition one-forms $\Lambda_{\alpha\beta}$. Note also that the gauge transformation of the $B$ field $\delta B_\alpha=d\imath_X B_\alpha$ turns the world-volume gauge connection back to the initial $A_\alpha$ with undeformed transition functions $h_{\alpha\beta}$. Then, supplemented by the gauge transformation, in this form the diffeomorfism generated by $X$ acts only on the $B$ field (and consequently on $H$) accordingly to the rule $\delta_X B=\imath_X H$, leaving all the transition functions and the world-volume gauge field untouched.   The resulting infinitesimal deformation of the world-volume field strength is given by $\delta\calf=P_\Sigma[\imath_X H]$.


\end{appendix}



\end{document}